\documentclass[english,reprint,aps,amsmath,amssymb,prl,superscriptaddress,floatfix]{revtex4-2}
\usepackage[T1]{fontenc}
\usepackage{color}
\usepackage{amsmath}
\usepackage{amssymb}
\usepackage{graphicx}

\makeatletter

\newcommand{\lyxmathsym}[1]{\ifmmode\begingroup\def\b@ld{bold}
  \text{\ifx\math@version\b@ld\bfseries\fi#1}\endgroup\else#1\fi}

\providecommand{\tabularnewline}{\\}

\usepackage{dcolumn}
\usepackage{bm}% bold mathsp
\usepackage{mathrsfs}
\usepackage[colorlinks,linkcolor=red,anchorcolor=blue,citecolor=green]{hyperref}
\allowdisplaybreaks[4]

\usepackage{amssymb}
\makeatletter

\newcommand{\Rmnum}[1]{\expandafter\@slowromancap\romannumeral #1@}
\makeatother

\makeatother

\usepackage{babel}
\begin{document}
\title{Electronic Hall viscosity: hidden indicator for antiferromagnets }
\author{Ding Li}
\affiliation{Anhui Key Laboratory of Low-Energy Quantum Materials and Devices,
High Magnetic Field Laboratory, HFIPS, Chinese Academy of Sciences,
Hefei, Anhui 230031, China}
\affiliation{Department of Physics, University of Science and Technology of China,
Hefei 230026, P.R. China}
\author{Tao Qin}
\affiliation{School of Physics, Anhui University, Hefei, Anhui Province 230601,
People’s Republic of China}
\author{Jianhui Zhou}
\email{jhzhou@hmfl.ac.cn}

\affiliation{Anhui Key Laboratory of Low-Energy Quantum Materials and Devices,
High Magnetic Field Laboratory, HFIPS, Chinese Academy of Sciences,
Hefei, Anhui 230031, China}
%\date{\today}
\begin{abstract}
The antiferromagnets with negligible stray fields and ultrafast spin
dynamics play a crucial role in the fields of energy-efficient spintronics
and topological electronics. However, the detection and control of
the underlying nontrivial Berry curvature become extremely limited
by the vanishing magnetization and anomalous Hall conductivity. Here,
we show the electronic Hall viscosity is closely related to the quadruple
Berry curvature of Bloch bands and is bounded by the $d$-orbit factor
modulated second moment of the quantum volume. Moreover, we derive
the symmetry requirement for nonzero electronic Hall viscosity that
could characterize antiferromagnetic ordering even when the linear
anomalous Hall response gets forbidden. We further examine our key
findings in two archetypal antiferromagnets: $d$-wave altermagnet
$\mathrm{RuO}_{2}$, and noncollinear $\mathrm{Mn_{3}Sn}$ through
direct first-principle calculations. Thus, our work reveals a new
and fundamental quantum geometry quantity of generic antiferromagnets
and offers a broadly applicable way to design antiferromagnetic spintronics
devices via unconventional Hall viscosity. 
\end{abstract}
\maketitle
\textit{\textcolor{black}{Introduction.-{}-}}Antiferromagnetism, one
of the most fundamental magnetic ordering, features zero net magnetization
\citep{Neel1971Science}. It could be realized in a plenty of compounds
and artificial structures and usually be categorized into several
basic classes according to the geometric configuration of spin moments,
including collinear, noncollinear, coplanar, noncoplanar classes.
Due to the advantages of zero stray field and ultra fast spin dynamics
(THz), antiferromagnets play a vital role in energy-efficient spintronics
\citep{Baltz2018RMP,Jungwirth2016NN} and fascinating topological
phases of matters \citep{Nagaosa2010RMP,Tokura2019NRP} and strongly
correlated physics \citep{AuerbachMag}. However, the control and
detection of the underlying Berry curvature and related quantum geometry
quantities of Bloch states \citep{Provost1980CMP,MaYQ2010PRB} turns
out to be extremely limited therein due to the vanishing anomalous
Hall conductivity within the linear response regime. 

The electronic Hall viscosity (EHV), one novel dispersionless transverse
response, is responsible for the strains of lattice and uncovers the
intrinsic quantum geometry properties of the charge and lattice degree
of freedom \citep{Avron1995PRL}, resembling the Chern number \citep{XiaoD2010RMP}.
Previous studies of the EHV mainly focus on the integer/fractional
quantum Hall systems \citep{Avron1995PRL,Tokatly2007PRB,Haldane2011PRL,Read2011PRB}
and some simple and ideal model systems of topological insulators
and topological superconductivity \citep{Hughes2011PRL,Shapourian2015PRB,Rao2020PRX}.
Moreover, the Hall viscosity in electronic liquids breaking time reversal
symmetry is expressed as the second derivative with respect to the
wave vector of generic dynamical anomalous Hall conductivity (AHC)
\citep{Hoyos2012PRL,Kozii2021PRL}, suggesting its strong dependence
of the finite AHC. However, how the magnetic ordering of spin affect
the EHV in the large amount of various magnetic materials remains
largely unexplored. Moreover, what are the quantum geometric nature
and symmetry properties of the EHV and how to quantitatively calculate
the EHV in realistic materials are still elusive. Whether EHV can
be disentangled with AHC and exist independently? 

In this work, we establish the intimate relation between the EHV and
the quadrupole moment of $\boldsymbol{k}$-space Berry curvature in
solids, estimate its quantum-geometry bound and determine the rule
for nonzero EHV. We derive rigorous and comprehensive symmetry conditions
for nonzero EHV in generic antiferromagnets. Through direct first-principles
calculations, we validate our theory in two representative antiferromagnets:
$\mathrm{RuO}_{2}$ and \emph{$\mathrm{Mn}_{3}\mathrm{Sn}$.} In addition,
we briefly discuss the experimental detections of EHV. 

\textit{\textcolor{black}{Formalism of EHV.-{}-}}In order to describe
the response of electronic liquid to deformation of crystal (such
as acoustic phonons), we start from the retarded force-force response
function $\chi_{\alpha\beta}(\boldsymbol{q},\omega)$ as \citep{Barkeshli2012PRB,LiuDH2017PRL,LiD2025arXiv}
\begin{equation}
\chi_{\alpha\beta}(\boldsymbol{q},\omega)=\hbar\sum_{n,m}\int_{\mathrm{BZ}}\left[d\boldsymbol{k}\right]F_{mn}(\omega,\boldsymbol{k},\boldsymbol{q})S_{nm}^{\alpha\beta}(\boldsymbol{k},\boldsymbol{q})\label{FFC}
\end{equation}
where the dynamical factor is $F_{mn}(\omega,\boldsymbol{k},\boldsymbol{q})=\left[f_{n}(\boldsymbol{k})-f_{m}(\boldsymbol{k}+\boldsymbol{q})\right]/\left[\hbar\omega+i\delta+\varepsilon_{nm}(\boldsymbol{k},\boldsymbol{q})\right]$
with $\varepsilon_{nm}(\boldsymbol{k},\boldsymbol{q})=\varepsilon_{n}(\boldsymbol{k})-\varepsilon_{m}(\boldsymbol{k}+\boldsymbol{q})$
being the difference of band energies and $f_{n}(\boldsymbol{k})$
being the Fermi distribution function. $\delta$ is a positive infinitesimal.
The subscript $\mathrm{BZ}$ is abbreviation for Brillouin zone. We
also used the notation of $\left[d\boldsymbol{k}\right]=\frac{d\boldsymbol{k}}{(2\pi)^{3}}$.
We adopt the electron-phonon interaction $\hat{\mathcal{H}}_{e-ph}$
to mimic the impact of deformation of crystal on electron states.
$S_{nm}^{\alpha\beta}(\omega,\boldsymbol{k},\boldsymbol{q})=\left\langle \psi_{n}(\boldsymbol{k})|\hat{T}_{\alpha}(-\boldsymbol{q})|\psi_{m}(\boldsymbol{k}')\right\rangle \left\langle \psi_{m}(\boldsymbol{k}')|\hat{T_{\beta}}(\boldsymbol{q})|\psi_{n}(\boldsymbol{k})\right\rangle $
is the product of the matrix elements of the force operator $\hat{T}_{\alpha}(\boldsymbol{q})=-\partial\hat{\mathcal{H}}_{e-ph}/\partial u_{\boldsymbol{q}\alpha}$
and thus quadratic in the electron-phonon interaction, where $u_{\boldsymbol{q}\alpha}$
is the Fourier transform of the displacement field along the $\alpha$-th
direction $u_{\alpha}(\boldsymbol{r})$. 
\begin{figure}
\includegraphics[width=8.4cm]{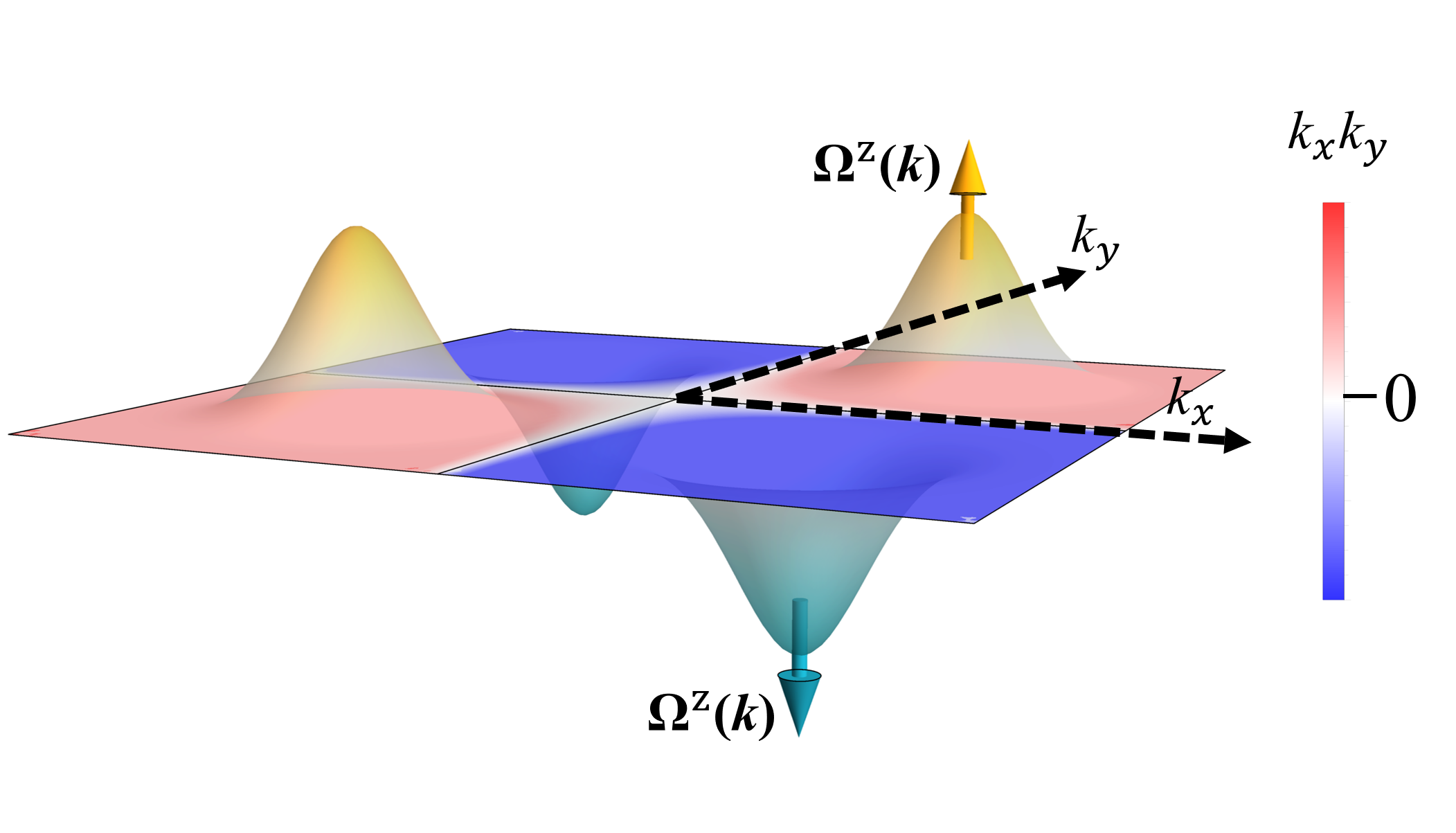}

\caption{Schematic of the electronic Hall viscosity $\eta_{xyz}=k_{x}k_{y}\Omega^{z}(\boldsymbol{k})$
for systems with $\mathcal{C}_{4}$ symmetry on the $k_{x}$-$k_{y}$
plane. The $d$-orbital function $k_{x}k_{y}$ is plotted on the plane.
The colored peaks denote the Berry curvature $\Omega^{z}(\boldsymbol{k})$
with the up/down arrow for the positive/negative sign, implying zero
anomalous Hall conductivity. \label{figEta}}
\end{figure}

For the long-wavelength deformations, the leading order EHV can be written as:
\begin{align}
\eta_{\gamma\delta\alpha\beta} & =\underset{\omega\rightarrow0}{\mathrm{lim}}\underset{\boldsymbol{q}\rightarrow0}{\mathrm{lim}}\frac{\partial}{\partial\omega}\frac{\partial^{2}}{\partial q_{\gamma}q_{\delta}}\mathrm{Im}\left[\chi_{\alpha\beta}^{H}(\boldsymbol{q},\omega)\right]
\end{align}
where $\chi_{\alpha\beta}^{H}(\boldsymbol{q},\omega)=\left(\chi_{\alpha\beta}(\boldsymbol{q},\omega)+\chi_{\beta\alpha}^{*}(\boldsymbol{q},\omega)\right)/2$
is the Hermitian part of $\chi_{\alpha\beta}(\boldsymbol{q},\omega)$
and suggests its dissipationless nature. In order to capture both
the local stretch and rotation of deformations \citep{EPhC}, we utilize
the tetrad field $e_{j}^{\mu}=\delta_{j}^{\mu}-\partial u_{j}(r)/\partial r_{\mu}$
\citep{Hughes2011PRL,Shapourian2015PRB,LiuDH2017PRL}, which has been
applied to the field theory in curved space and the effective theory
of crystals in the presence of strains and topological defects \citep{Nakahara2003GTP}.
Accordingly, the EHV tensor takes the compact form at zero temperature
\begin{align}
\eta_{\gamma\delta\alpha\beta} & =-\hbar\int_{\mathrm{BZ}}\left[d\boldsymbol{k}\right]k_{\gamma}k_{\delta}\Omega_{\alpha\beta}\left(\boldsymbol{k}\right)\label{eq:etaBC}
\end{align}
where $\Omega_{\alpha\beta}(\boldsymbol{k})=\sum_{\varepsilon_{n}\left(\boldsymbol{k}\right)<\epsilon_{F}}\Omega_{n,\alpha\beta}(\boldsymbol{k})$
with $\Omega_{n,\alpha\beta}(\boldsymbol{k})=-2\mathrm{Im}\sum_{m\neq n}\frac{v_{nm}^{\alpha}(\boldsymbol{k})v_{mn}^{\beta}(\boldsymbol{k})}{\left(\varepsilon_{n}\left(\boldsymbol{k}\right)-\varepsilon_{m}\left(\boldsymbol{k}\right)\right)^{2}}$
being the Berry curvature of the $n$-th energy band and $\epsilon_{F}$
being the Fermi energy \citep{XiaoD2010RMP}. It is convenient to
use the vector of Berry curvature $\Omega^{\lambda}(\boldsymbol{k})=\varepsilon_{\lambda\alpha\beta}\Omega_{\alpha\beta}(\boldsymbol{k})$
with $\varepsilon_{\lambda\alpha\beta}$ being the Levi-Civita symbol.
Then the corresponding EHV becomes $\eta_{\gamma\delta\lambda}\equiv\varepsilon_{\lambda\alpha\beta}\eta_{\gamma\delta\alpha\beta}$.
Fig. (\ref{figEta}) shows the primary physics of $\eta_{xyz}$ for
$\mathcal{C}_{4}$ symmetric systems. 

Several remarks are in order here. First, Eq. (\ref{eq:etaBC}) reveals
that the EHV directly corresponds to the integration of the quadrupole
moment of the Berry curvature in momentum space, similar to the AHC
as an integration of Berry curvature \citep{Nagaosa2010RMP}. This
establishes a profound connection between the quantum geometry of
electronic states and the viscoelastic response function. Second,
the adiabatic curvature in Eq. $\left(\ref{FFC}\right)$ usually acts
as the molecular Berry curvature for the chiral phonons and the phonon
magnetic moment in insulating solids \citep{ZhangLF2015PRL,Saparov2022PRB,ZhangTT2022PRR,ChenYR2025PRL,chatterjee2026arxiv,ChenHR2026arxiv}.
Third, our linear response derivation based on the Bloch states is
quite general and does not require the fine symmetrization of Belinfante-Rosenfeld
stress tensor in two-dimensional electron liquids \citep{Rao2020PRX}. 

The quantum geometry of Bloch states sets the fundamental bound on
the a variety of physical response functions \citep{Onishi2024PRX,Pai2026PRL}.
In order to derive the bound for EHV $\eta_{\gamma\delta\lambda}$
above, we need separate $\eta_{\gamma\delta\lambda}$ into the contribution
from the $\boldsymbol{k}$-space $d$-orbital function and the Berry
curvature. Hence, we apply the Cauchy-Schwarz inequality for any bounded
and summable functions \citep{CSIneq} to the square of EHV $\eta_{\gamma\delta\lambda}^{2}$
and have 
\begin{align}
 & \left(\int_{\mathrm{BZ}}\left[d\boldsymbol{k}\right]k_{\gamma}k_{\delta}\Omega^{\lambda}\left(\boldsymbol{k}\right)\right)^{2}\nonumber \\
\leq & \left(\int_{\mathrm{BZ}}\left[d\boldsymbol{k}\right]k_{\gamma}^{2}k_{\delta}^{2}\right)\times\left(\int_{\mathrm{BZ}}\left[d\boldsymbol{k}\right]\left[\Omega^{\lambda}\left(\boldsymbol{k}\right)\right]^{2}\right)
\end{align}
The integral of square of $d$-orbital function over the Brillouin
zone in the parentheses gives rise to an orbital weight factor or
volume $\mathrm{v}_{d}$. 

Let us estimate the bound of the integration of square of the Berry
curvature. First, from the semipositivity of the quantum geometry
tensor $\mathcal{Q}\left(\boldsymbol{k}\right)=g\left(\boldsymbol{k}\right)+i\Omega\left(\boldsymbol{k}\right)/2$,
that is $\mathbf{\mathrm{det}}\mathcal{Q}\left(\boldsymbol{k}\right)\geq0$,
we could have the basic inequality between the Berry curvature and
the quantum metric $\mathbf{\mathrm{det}}\left(g\right)\geq\left|\Omega\right|^{2}/4$
\citep{Roy2014PRB}, immediately leading to the relation of $\int_{\mathrm{BZ}}\left[d\boldsymbol{k}\right]\left[\Omega^{\lambda}\left(\boldsymbol{k}\right)\right]^{2}\leq4\int_{\mathrm{BZ}}\left[d\boldsymbol{k}\right]\mathbf{\mathrm{det}}\left(g\right).$
It is known that, in two-dimensional case, the dimensionless quantum
volume of the Brillouin zone is defined as $\mathrm{vol}_{g}\equiv\int_{\mathrm{BZ}}\left[d\boldsymbol{k}\right]\sqrt{\mathbf{\mathrm{det}}\left(g\right)}$
\citep{Ozawa2021PRB}. According to the dimensional analysis, the
dimension of $\mathbf{\mathrm{det}}\left(g\right)$ is $k^{-4}$.
In analogy, the integral of $\mathbf{\mathrm{det}}\left(g\right)$
can be regarded as the second moment of the quantum volume as $\mathrm{vol}_{g}^{2nd}\equiv\int_{\mathrm{BZ}}\left[d\boldsymbol{k}\right]\mathbf{\mathrm{det}}\left(g\right)$.
By collecting all these results, we thus obtain the fundamental upper
bound of the EHV as 
\begin{equation}
\eta_{\gamma\delta\lambda}\leq2\hbar\sqrt{\mathrm{v}_{d}\cdot\mathrm{vol}_{g}^{2nd}}.
\end{equation}
Remarkably, it is clear that the EHV is bounded by the $d$-orbit
factor modulated second moment of the quantum volume. It is a nontrivial
and distinct multipole counterpart of the classical inequality between
the the quantum volume and the Chern number $\mathcal{C}$ of $\mathrm{vol}_{g}\geq\pi\left|\mathcal{C}\right|$
\citep{Julku2016PRL}. In fact, the inequality can be applicable to
the other higher-order moments of Berry curvature. The main modification
is to replace $\mathrm{v}_{d}$ with the higher-orbit one. It is one
of the key results of this work. 

\textit{\textcolor{black}{Multipole expansion of Berry curvature.-{}-}}At
first, we brief their behavior under discrete symmetries. Specifically,
under time reversal ($\mathcal{T}$) and spatial inversion ($\mathcal{P}$),
the kernel $\tilde{\Omega}_{\gamma\delta\lambda}(\boldsymbol{k})\equiv k_{\gamma}k_{\delta}\Omega^{\lambda}$
transforms as $\mathcal{T}:\tilde{\Omega}_{\gamma\delta\lambda}(\boldsymbol{k})\rightarrow-\tilde{\Omega}_{\gamma\delta\lambda}(-\boldsymbol{k})$
and $\mathcal{P}:\tilde{\Omega}_{\gamma\delta\lambda}(\boldsymbol{k})\rightarrow\tilde{\Omega}_{\gamma\delta\lambda}(-\boldsymbol{k})$,
leading to 
\begin{align}
\mathcal{PT} & :\tilde{\Omega}_{\gamma\delta\lambda}(\boldsymbol{k})\rightarrow-\tilde{\Omega}_{\gamma\delta\lambda}(\boldsymbol{k}).\label{PTOmega}
\end{align}
It has the same transformation properties as Berry curvature $\Omega^{\lambda}(\boldsymbol{k})$.
Consequently, $\mathcal{PT}$ symmetry enforces both EHV and AHC vanish.
Next, we shall focus on the role of rotational symmetries, such as
$n$-fold rotation about $\nu$-direction $\mathcal{C}_{n\nu}$ and
their combinations with $\mathcal{T}$, $\mathcal{P}$ and $\mathcal{TP}$
($\mathcal{P}\mathcal{C}_{n\nu}$,$\mathcal{T}\mathcal{C}_{n\nu}$,
or $\mathcal{PT}\mathcal{C}_{n\nu}$) on the EHV. 

In order to determine the properties of quadrupole (also multipole)
Berry curvature under the rotational symmetries, it is instructive
to utilize the spherical harmonics in momentum space, in analogy to
the multipole expansion in classical electromagnetism and $f$-electron
materials \citep{Santini2009RMP}. First, the weighting factors $k_{\gamma}k_{\delta}$,
transforming as rank-2 spherical tensors, can be decomposed into the
spherical harmonics $\boldsymbol{Y}_{2}^{m}$ with the magnetic quantum
numbers $m=0,\pm1,\pm2$ as 
\begin{equation}
k_{\gamma}k_{\delta}=k^{2}\sum_{m}d_{m}\boldsymbol{Y}_{l}^{m}(\theta,\phi),
\end{equation}
where $d_{m}$ are some coefficients, $\theta$ and $\phi$ are the
spherical coordinates for the vector $\boldsymbol{k}$ \citep{Sakurai2020MQM}.
Second, the Berry curvature $\Omega^{\lambda}(\boldsymbol{k})$ can
be also expanded in terms of the $k$-space spherical harmonics $\boldsymbol{Y}_{l}^{m}$:
\begin{equation}
\Omega^{\lambda}(\boldsymbol{k})=\sum_{lm}c_{lm}^{\lambda}(k)\boldsymbol{Y}_{l}^{m}(\theta,\phi),
\end{equation}
where the coefficients $c_{lm}^{\lambda}$ reflect the inner geometric
structure of the Berry curvature in momentum space. Owing to the orthogonality
relation $\left\langle l'm'|lm\right\rangle =\delta_{l'l}\delta_{m'm}$,
if the expansion contains the $d$-wave like spherical harmonics $\boldsymbol{Y}_{2}^{m}$,
the corresponding integration should be non-zero in principle. As
a result, the EHV reduces to be 
\begin{align}
\eta_{\gamma\delta\lambda}^ {} & =-\frac{\hbar}{(2\pi)^{3}}\sum_{m}d_{m}\int dkc_{2m}^{\lambda}\left(k\right)k^{4}.
\end{align}
The general procedures above could be straightforwardly generalized
to the multipole Berry curvature and orbital/heat magnetic moment
\citep{Tahir2023PRL,LiD2025arXiv}. After some complicated manipulations,
we list the expansion of $\Omega^{\alpha}(\boldsymbol{k})$ under
all rotational and the combined symmetry operations along the $\alpha$-axis
with $\alpha=x,y,z$ in the Supplemental Material (SM) \citep{SMs}.

It is known that the AHC component vanishes when the Berry curvature
satisfies the condition $\Omega^{\lambda}\left(R\boldsymbol{k}\right)=-\text{\ensuremath{\Omega^{\lambda}}}\left(\boldsymbol{k}\right)$,
where $R$ is an element of the group of $\boldsymbol{k}$ \citep{Suzuki2017PRB,CaoJ2023PRL,Kurumaji2023PRR}.
For instance, under $\mathcal{T}\mathcal{C}_{4z}$ symmetry, the Berry
curvature satisfies $\Omega^{z}(\mathcal{C}_{4z}\boldsymbol{k})=-\Omega^{z}(\boldsymbol{k})$,
leading to $\sigma^{z}=0$. By contrast, the quadrupolar component
with $m=\pm1$ (namely $k_{x}k_{y}$ and $k_{x}^{2}-k_{y}^{2}$) acquires
an additional minus sign under the $\mathcal{T}\mathcal{C}_{4z}$
rotation. Consequently, it will compensate negative sign before the
transformed Berry curvature, yielding finite EHV components $\eta_{xxz}$,
$\eta_{xyz}$ (See Fig. (\ref{figEta})) and $\eta_{yyz}$ even when
the AHC is strictly forbidden by symmetry. It is another key result
of this work. 

We summarize the transformation properties of $\Omega^{z}(\boldsymbol{k})$
and the irreducible representation decomposition of $k_{\gamma}k_{\delta}$
under the symmetry operations $\mathcal{R}$ in Table \ref{tab:SYM-EHV}.
We find that both $\mathcal{C}_{nz}$ and $\mathcal{P}\mathcal{C}_{nz}$
preserve the sign of the Berry curvature, thereby permitting a non-zero
AHC. Consequently, a non-zero quadrupole moment arises only for those
$k_{\gamma}k_{\delta}$ components that transform with a factor of
$+1$ under $\mathcal{C}_{nz}$ or $\mathcal{P}\mathcal{C}_{nz}$.
\begin{table*}
\caption{\label{tab:SYM-EHV} The second to fifth rows are the transformation
properties of the Berry curvature $\Omega^{z}(\boldsymbol{k})$ and
$k_{\gamma}k_{\delta}$ under the operators $\mathcal{R}$, where
$\mathcal{R}$ denotes $\mathcal{C}_{nz}$, $\mathcal{P}\mathcal{C}_{nz}$,
$\mathcal{T}\mathcal{C}_{nz}$, and $\mathcal{PT}\mathcal{C}_{nz}$.
The last two rows are the permitted non-zero AHC and EHV tensor components.}

\begin{ruledtabular}
\begin{tabular}{c||cccccccc}
\multicolumn{2}{c}{} & $\mathcal{C}_{2z}$,$\mathcal{P}\mathcal{C}_{2z}$ & $\mathcal{C}_{3z}$,$\mathcal{P}\mathcal{C}_{3z}$ & $\mathcal{C}_{4z}$,$\mathcal{P}\mathcal{C}_{4z}$ & $\mathcal{C}_{6z}$,$\mathcal{P}\mathcal{C}_{6z}$ & $\mathcal{C}_{2z}'$,$\mathcal{P}\mathcal{C}_{2z}'$ & $\mathcal{C}_{4z}'$,$\mathcal{P}\mathcal{C}_{4z}'$ & $\mathcal{C}_{6z}'$,$\mathcal{P}\mathcal{C}_{6z}'$\tabularnewline
\hline 
\multicolumn{2}{c}{$\Omega^{z}(\boldsymbol{k})$} & $\Omega^{z}(\mathcal{R}^{-1}\boldsymbol{k})$ & $\Omega^{z}(\mathcal{R}^{-1}\boldsymbol{k})$ & $\Omega^{z}(\mathcal{R}^{-1}\boldsymbol{k})$ & $\Omega^{z}(\mathcal{R}^{-1}\boldsymbol{k})$ & –$\Omega^{z}(\mathcal{R}^{-1}\boldsymbol{k})$ & –$\Omega^{z}(\mathcal{R}^{-1}\boldsymbol{k})$ & –$\Omega^{z}(\mathcal{R}^{-1}\boldsymbol{k})$\tabularnewline
\multicolumn{2}{c}{$2k_{z}^{2}-k_{x}^{2}-k_{y}^{2}$} & +1 & +1 & +1 & +1 & +1 & +1 & +1\tabularnewline
\multicolumn{2}{c}{$k_{x}k_{z}$,$k_{y}k_{z}$} & -1 & $e^{\mp i2\pi/3}$ & $\mp i$ & $e^{\mp i\pi/3}$ & -1 & $\mp i$ & $e^{\mp i\pi/3}$\tabularnewline
\multicolumn{2}{c}{$k_{x}^{2}-k_{y}^{2}$, 2$k_{x}k_{y}$} & +1 & $e^{\pm i2\pi/3}$ & -1 & $e^{\mp i2\pi/3}$ & +1 & -1 & $e^{\mp i2\pi/3}$\tabularnewline
\multicolumn{2}{c}{AHC} & $\sigma^{z}\neq0$ & $\sigma^{z}\neq0$ & $\sigma^{z}\neq0$ & $\sigma^{z}\neq0$ & $\sigma^{z}=0$ & $\sigma^{z}=0$ & $\sigma^{z}=0$\tabularnewline
\multicolumn{2}{c}{EHV} & $\eta_{xxz}$,$\eta_{yyz}$,$\eta_{xyz}$,$\eta_{zzz}$ & $\eta_{xxz}$,$\eta_{yyz}$,$\eta_{zzz}$ & $\eta_{xxz}$,$\eta_{yyz}$,$\eta_{zzz}$ & $\eta_{xxz}$,$\eta_{yyz}$,$\eta_{zzz}$ & $\eta_{xzz}$,$\eta_{yzz}$ & $\eta_{xxz}$,$\eta_{xyz}$,$\eta_{yyz}$ & $\eta_{xxz}$,$\eta_{yyz}$,$\eta_{zzz}$\tabularnewline
\end{tabular}
\end{ruledtabular}

\end{table*}
 Next we apply our theory above to general magnetic crystal symmetries
and specific materials, further determine the non-vanishing EHV and
AHC components. 

\emph{Symmetry of EHV}\textit{\textcolor{black}{.-{}-}}Microscopic
symmetries play a vital role in the structure of response tensors
to various external perturbations \citep{Kleiner1966PR,CaoJ2023PRL,Kurumaji2023PRR,ZhouJD2022Nature}.
Let us determine the symmetry properties of EHV in generic antiferromagnets.
Due to the kernel $\tilde{\Omega}_{\gamma\delta\lambda}(\boldsymbol{k})$
is translationally invariant, only the magnetic point group (MPG)
imposes constraints on the form of EHV \citep{Kleiner1966PR}. First,
according to Eq. $\left(\ref{PTOmega}\right)$, the EHV is absent
in both $\mathcal{PT}$ and $\tau\mathcal{T}$-symmetric antiferromagnets,
where $\tau\mathcal{T}$ represents a half-lattice translation ($\tau$)
combined with $\mathcal{T}$. Next, we mainly consider the symmetry
condition of nonzero EHV that is constrained by a composite symmetry
of either $\mathcal{T}\mathcal{C}_{n\nu}$ or $\mathcal{PT}\mathcal{C}_{n\nu}$. 

Under the MPG operation $\hat{\mathcal{O}}=\mathcal{T}^{s_{\mathcal{T}}}\hat{\mathcal{R}}$,
where $\hat{\mathcal{R}}$ is a spatial operation, $s_{\mathcal{T}}=1(0)$
for operations with (no) $\mathcal{T}$, and $|\mathcal{R}|$=det($\mathcal{R}$),
the EHV tensor transforms as: $\eta_{\gamma\delta\lambda}\rightarrow\eta_{\gamma'\delta'\lambda'}=(-1)^{s_{\mathcal{T}}}|\mathcal{R}|\mathcal{R}_{\gamma'\gamma}\mathcal{R}_{\delta'\delta}\mathcal{R}_{\lambda'\lambda}\eta_{\gamma\delta\lambda}$.
From this transformation rule, one can see that time reversal $\mathcal{T}$
(present when $s_{\mathcal{T}}=1$) simply flips the sign of all components.
Notably, spatial inversion $\mathcal{P}$ (for which $|\mathcal{R}|=-1$)
imposes no independent constraint on $\eta_{\gamma\delta\lambda}$,
because its sign change from the determinant is compensated by the
fact that $\mathcal{R}_{\gamma'\gamma}\mathcal{R}_{\delta'\delta}\mathcal{R}_{\lambda'\lambda}=-1$
for a pure inversion. 

Thus, we can restrict our classification to magnetic Laue groups (MLGs)
\citep{Seemann2015PRB}, in which we treat spatial inversion and mirror
planes as an identity operation and two-fold rotations about axes
normal to the planes, respectively. We have listed the allowed non-vanishing
components of the EHV tensor $\eta_{\gamma\delta\lambda}$ and the
AHC $\sigma^{\alpha}$ for all 10 MLGs describing antiferromagnets
breaking both $\mathcal{PT}$ and $\tau\mathcal{T}$ symmetry in SM
\citep{SMs}. 

We find that the structure of the EHV tensor $\eta_{\gamma\delta\lambda}$
is exclusively constrained by the $\mathcal{C}_{n\nu}^{\prime}$ operators
present within the MLG, where $\mathcal{C}_{n\nu}^{\prime}\equiv\mathcal{T}\mathcal{C}_{n\nu}$
denotes an $n$-fold rotation ($n=2,4,6$) about the $\nu$-axis combined
with time reversal symmetry. Explicitly, the constraints imposed by
these composite symmetries are as follows: (I) $\mathcal{C}_{2\nu}^{\prime}$
symmetry allows nonzero components $\eta_{\gamma\delta\lambda}$ if
the $\nu$-oriented index appears twice in the set of $\{\gamma,\delta,\lambda\}$.
(II) $\mathcal{C}_{4\nu}'$ symmetry forces $\eta_{\gamma\delta\lambda}$
to vanish unless the set $\{\gamma,\delta,\lambda\}$ contains exactly
one $\nu$-index. (III) $\mathcal{C}_{6\nu}^{\prime}$ symmetry forces
all components $\eta_{\gamma\delta\lambda}$ to vanish once the $\nu$-index
appears in $\{\gamma,\delta,\lambda\}$. 

For example, for the MPG $m'm'2$, the corresponding MLG is $m'm'm$
and contains three symmetry operators: $\mathcal{C}_{2z}$, $\mathcal{C}_{2x}'$
and $\mathcal{C}_{2y}'$. The tensor structure is constrained by the
two time-reversal-incorporated rotations, $\mathcal{C}_{2x}^{\prime}$
and $\mathcal{C}_{2y}^{\prime}$. According to the rule for $\mathcal{C}_{2\nu}'$
above, they force the vanishing of any component where the number
of $x$-indices or the number of $y$-indices is odd. Applying these
constraints leaves only the following non-zero independent components:
$\eta_{xxz}$, $\eta_{xzx}$, $\eta_{yyz}$, $\eta_{yzy}$ and $\eta_{zzz}$.
Next, we shall apply our theory to two representative antiferromagnets. 

\begin{figure}
\includegraphics[width=8.5cm]{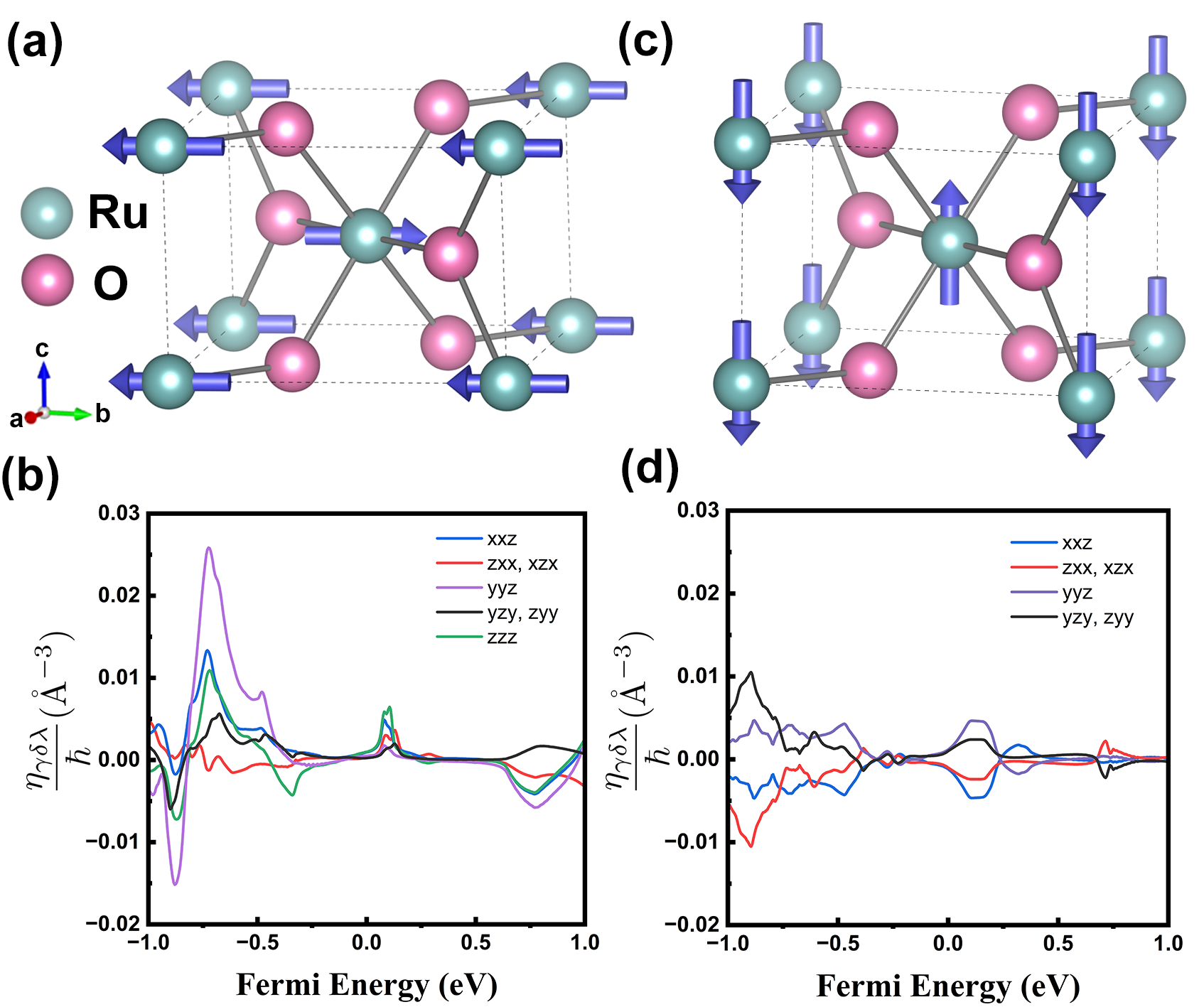}

\caption{\label{fig:EHVRuO} Two different magnetic configurations of the altermagnet
$\mathrm{RuO}_{2}$\emph{ }with Néel vector $\boldsymbol{L}$ along
(a) $a$-axis and (c) $c$-axis. (b) and (d) The corresponding EHV
tensor versus the Fermi level.}
\end{figure}

\emph{d-wave altermagnet $RuO_{2}$}\textit{\textcolor{black}{.-{}-}}Recently,
altermagnets have garnered considerable attention owing to their unique
alternating spin splitting in momentum space \citep{Sandratskii1981PSSb,WuCJ2007PRB,YuanLD2020PRB,Smejkal2021SA,MaHY2021NC,Smejkal2022PRX3,Mazin2022PRX},
giving rise to a variety of novel physical phenomena \citep{BaiL2024AFM,SongC2025NRM,Jungwirth2025Newton}.
Numerous altermagnetic materials have been theoretically predicted
\citep{LiuPF2022PRX,XiaoZY2024PRX,XiaoRC2026SCPMA}, amongst which
$\mathrm{RuO}_{2}$ is one of the intensively investigated and debatable
compounds \citep{Smejkal2021SA,BaiH2022PRL,Hiraishi2024PRL,Fedchenko2024SA,Chen2026NN}.
$\mathrm{RuO}_{2}$ crystallises in space group $P4_{2}/mnm$ \citep{ZhouXD2024PRL},
and its magnetic structure is illustrated in Fig. \ref{fig:EHVRuO}(a)
and (c) for the Néel vector oriented along the a-axis and c-axis with
associated magnetic space groups $Pnn'm'$ and $P4_{2'}/mnm'$, respectively.
Specifically, when the Néel vector $\boldsymbol{L}$ lies along the
a-axis, $\mathrm{RuO}_{2}$ belongs to the MLG $m'm'm$, which permits
only $\sigma^{z}$ to be non-zero (see Table-S3 in SM \citep{SMs}).
However, when $\boldsymbol{L}$ points along the $c$-axis, $\mathrm{RuO}_{2}$
falls into the MLG $4'/mm'm$, and all of the components of AHC vanish
(see Table-S5 in SM). Hence, a new physical quantity is desired to
characterize the antiferromagnetic order when AHC is forbidden. 

Figs. \ref{fig:EHVRuO}(b) and (d) display the density functional
theory (DFT) calculations of the EHV tensors for the two magnetic
configurations shown in Fig. \ref{fig:EHVRuO}(a) and (c), respectively.
It is evident that as the Néel vector $\boldsymbol{L}$ rotates from
the $c$-direction to the a-direction, the magnitudes and signs of
the corresponding EHV components alter obviously. Meanwhile, a new
non-zero component $\eta_{zzz}$ emerges. For example, for the magnetic
configuration in Fig. \ref{fig:EHVRuO}(c), one finds $\eta_{zxx}=\eta_{xzx}$,
$\eta_{yzy}=\eta_{zyy}$, and $\eta_{xxz}=-\eta_{yyz}$. Interestingly,
flipping the direction of the Néel vector $\boldsymbol{L}\rightarrow-\boldsymbol{L}$,
the EHV would change its sign \citep{SMs}. In fact, the essential
physics could be captured by the effective two-band model for $d$-wave
like Fermi surface with Rashba spin-orbit coupling \citep{SMs}. As
a result, the EHV acts as a smoking-gun transport signature of the
unique spin-splitting Fermi surface in $d$-wave altermagnets, including
other similar compounds: $\mathrm{MnF_{2}}$ \citep{Bhowal2024PRX},
$\mathrm{Mn_{5}Si_{3}}$ \citep{Reichlova2024NC} and $\mathrm{KV_{2}Se_{2}O}$
\citep{JiangB2025NP}. 

\begin{figure}
\includegraphics[width=8.5cm]{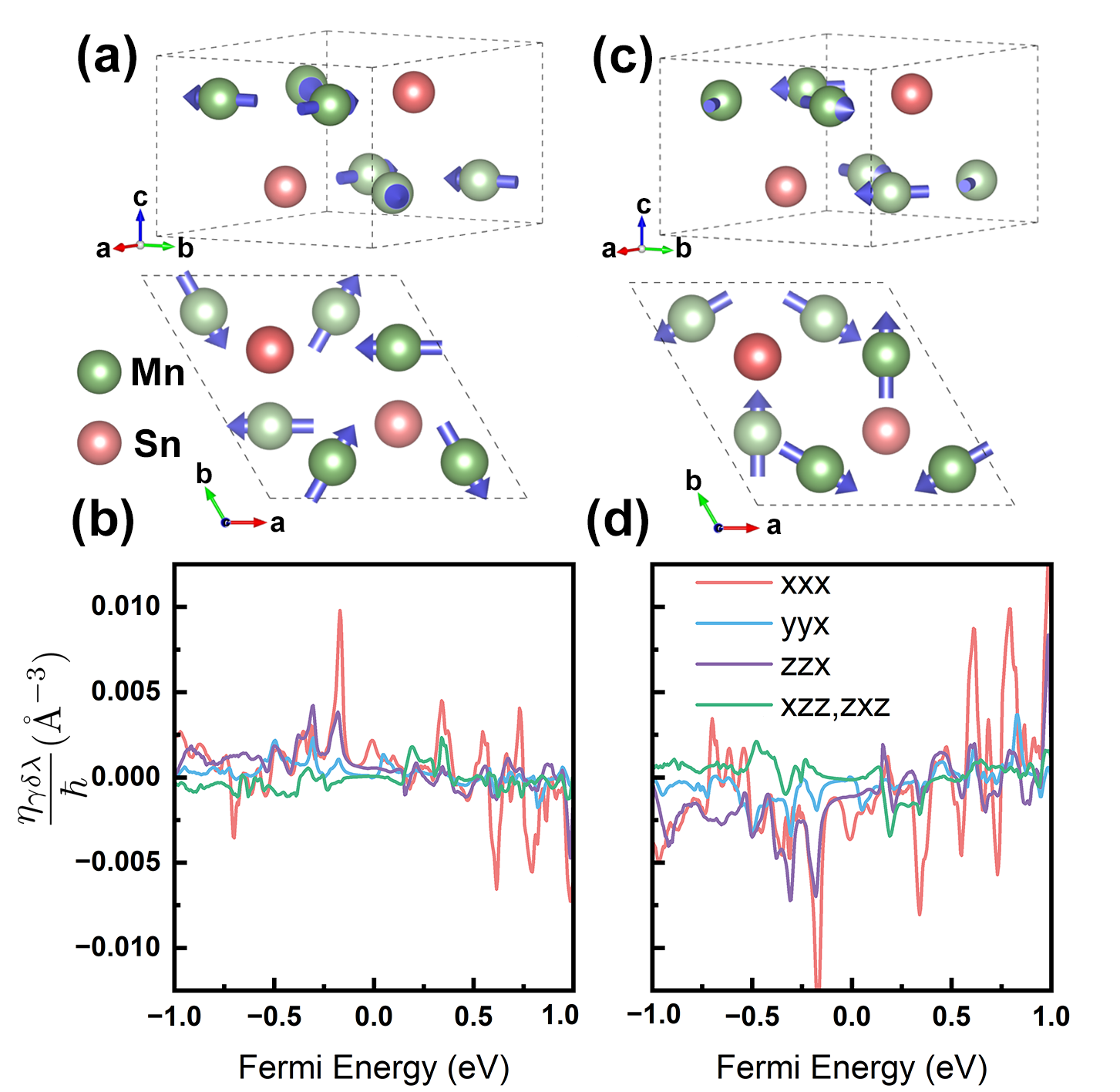}

\caption{\label{fig:EHVMnSn} Magnetic structure of the noncollinear antiferromagnets
\emph{$\mathrm{Mn}_{3}\mathrm{Sn}$} and the calculated EHV tensor.
(a) the type-III configuration of \emph{$\mathrm{Mn}_{3}\mathrm{Sn}$}
and (b) selected nonzero components of the EHV.\emph{ }(c) the type-IV
configuration of\emph{ $\mathrm{Mn}_{3}\mathrm{Sn}$} and (d) selected
nonzero components of the EHV. }
\end{figure}

\emph{Noncollinear antiferromagnet $Mn_{3}Sn$}\textit{\textcolor{black}{.-{}-}}\emph{$\mathrm{Mn}_{3}\mathrm{Sn}$}
family with coplanar spin order exhibits novel Hall transport phenomena
\citep{ChenH2014PRL,Nakatsuji2015Nature,LiuJP2017PRL,LiXK2017PRL}
and potential applications in antiferromagnetic spintronics \citep{Kimata2019Nature,Tsai2020Nature,Go2022PRL,Tsai2026Science}.
It crystallizes in a hexagonal structure with space group $P6_{3}/mmc$
(No.194) \citep{Rimmler2025NRM}. The Mn atoms constitute a two-dimensional
Kagome network, stacked in an A-B-A-B sequence along the $c$-axis
with Sn atoms occupying the interlayers. Below the Néel temperature
$\boldsymbol{T}_{N}=420$ K, the system orders into a noncollinear
antiferromagnetic state with propagation vector $\boldsymbol{k}=0$.
The related magnetic moments in \emph{$\mathrm{Mn}_{3}\mathrm{Sn}$}
shows a $120\lyxmathsym{\textdegree}$ arrangement within the $ab$-plane,
forming a triangular spiral structure accompanied by a weak ferromagnetic
order. Recent experiments suggest the two candidate configurations:
type-III and type-IV, depicted in Fig. \ref{fig:EHVMnSn}(a) and (c),
respectively \citep{cederholm2026arxiv}. Due to the nearly degenerate
calculated energies of both configurations \citep{Mn3Sn}, thus the
identification of the true ground state in experiments turns out to
be quite challenging. 

Both type-III and type-IV configurations belong to MPGs $2'/m'$ and
$m'$, respectively, share the same MLG $2'/m'$ and is connected
by a delicate rigid-body rotation \citep{LiuZ2026PRL}. Thus, both
magnetic configurations have the same structure of EHV and AHC (see
Table-S2 in SM). To be specific, the EHV tensor comprises 14 nonzero
components, of which 10 are independent. Meanwhile, the AHC satisfies
$\sigma^{x}\neq0$ and $\sigma^{z}\neq0$, whereas $\sigma^{y}=0$.
Figs. \ref{fig:EHVMnSn}(b) and (d) show the numerical results of
the partial EHV tensors for both magnetic configurations (Results
of other the nonzero components of the EHV are given in \citep{SMs}).
Remarkably, the magnitude and sign of some components of the EHV tensor
exhibit dramatic differences in the type-III and type-IV configurations,
which might stem from the distinct electronic states. Consequently,
the anisotropic EHV serves as a decisive probe for the true magnetic
ground state of \emph{$\mathrm{Mn}_{3}\mathrm{Sn}$}. 

Let us brief the experimental detection of the EHV in antiferromagnets.
First, for the metallic systems, the local voltages near the vicinity
of current-injecting contacts has been used to identify the Hall viscosity
of graphene's electron liquid in the presence of magnetic fields \citep{Berdyugin2019Science}.
The nonlocal Hall measurement could also detect the EHV that is responsible
for inhomogeneous electron fields or acoustic phonons \citep{Levitov2016NP,Kozii2021PRL}.
For the insulating case, the Hall viscosity in turn manifests in the
phonon physics and can be accessed by the acoustic Faraday effect
\citep{Shragai2026Nature}. 

\textit{\textcolor{black}{Conclusions.-{}-}}We found the intrinsic
relation between EHV and quadrupole of Berry curvature, revealed its
fundamental quantum-geometry upper bound and determined the condition
for finite EHV. Meanwhile, we systematically analyzed all possible
non-zero component of the EHV in generic antiferromagnets even with
vanishing AHC. We further verified our theory in two representative
antiferromagnetic materials via direct DFT calculations and briefly
discussed its experimental detection. 

Our work provides a novel and universal transport signature of the
antiferromagnetic order, facilitating their promising applications
in energy-efficient spintronics. In fact, our analysis based on the
$\boldsymbol{k}$-space spherical harmonics could be applicable to
the quantum phenomena due to multipole of other quantum geometric
quantities such as orbital/heat magnetic moments of Bloch electrons
in non/magnetic materials. 

The authors thank Guang-Yue Ji and Zheng Liu for useful discussions.
This work was financially supported by the National Key R\&D Program
of the MOST of China (Grant No. 2024YFA1611300), the National Natural
Science Foundation of China (Grant No. 12574059), HFIPS Director's
Fund (Grant No. BJPY2023B05), Anhui Provincial Major S\&T Project
(s202305a12020005) and the Basic Research Program of the Chinese Academy
of Sciences Based on Major Scientific Infrastructures (Grant No. JZHKYPT-2021-08)
and the High Magnetic Field Laboratory of Anhui Province under contract
No. AHHM-FX-2020-02.

\bibliographystyle{apsrev4-2}
\bibliography{EHV}

\begin{thebibliography}{87}%
\makeatletter
\providecommand \@ifxundefined [1]{%
 \@ifx{#1\undefined}
}%
\providecommand \@ifnum [1]{%
 \ifnum #1\expandafter \@firstoftwo
 \else \expandafter \@secondoftwo
 \fi
}%
\providecommand \@ifx [1]{%
 \ifx #1\expandafter \@firstoftwo
 \else \expandafter \@secondoftwo
 \fi
}%
\providecommand \natexlab [1]{#1}%
\providecommand \enquote  [1]{``#1''}%
\providecommand \bibnamefont  [1]{#1}%
\providecommand \bibfnamefont [1]{#1}%
\providecommand \citenamefont [1]{#1}%
\providecommand \href@noop [0]{\@secondoftwo}%
\providecommand \href [0]{\begingroup \@sanitize@url \@href}%
\providecommand \@href[1]{\@@startlink{#1}\@@href}%
\providecommand \@@href[1]{\endgroup#1\@@endlink}%
\providecommand \@sanitize@url [0]{\catcode `\\12\catcode `\$12\catcode
  `\&12\catcode `\#12\catcode `\^12\catcode `\_12\catcode `\%12\relax}%
\providecommand \@@startlink[1]{}%
\providecommand \@@endlink[0]{}%
\providecommand \url  [0]{\begingroup\@sanitize@url \@url }%
\providecommand \@url [1]{\endgroup\@href {#1}{\urlprefix }}%
\providecommand \urlprefix  [0]{URL }%
\providecommand \Eprint [0]{\href }%
\providecommand \doibase [0]{https://doi.org/}%
\providecommand \selectlanguage [0]{\@gobble}%
\providecommand \bibinfo  [0]{\@secondoftwo}%
\providecommand \bibfield  [0]{\@secondoftwo}%
\providecommand \translation [1]{[#1]}%
\providecommand \BibitemOpen [0]{}%
\providecommand \bibitemStop [0]{}%
\providecommand \bibitemNoStop [0]{.\EOS\space}%
\providecommand \EOS [0]{\spacefactor3000\relax}%
\providecommand \BibitemShut  [1]{\csname bibitem#1\endcsname}%
\let\auto@bib@innerbib\@empty
%</preamble>
\bibitem [{\citenamefont {Néel}(1971)}]{Neel1971Science}%
  \BibitemOpen
  \bibfield  {author} {\bibinfo {author} {\bibfnamefont {L.}~\bibnamefont
  {Néel}},\ }\href {https://doi.org/10.1126/science.174.4013.985} {\bibfield
  {journal} {\bibinfo  {journal} {Science}\ }\textbf {\bibinfo {volume}
  {174}},\ \bibinfo {pages} {985} (\bibinfo {year} {1971})}\BibitemShut
  {NoStop}%
\bibitem [{\citenamefont {Baltz}\ \emph {et~al.}(2018)\citenamefont {Baltz},
  \citenamefont {Manchon}, \citenamefont {Tsoi}, \citenamefont {Moriyama},
  \citenamefont {Ono},\ and\ \citenamefont {Tserkovnyak}}]{Baltz2018RMP}%
  \BibitemOpen
  \bibfield  {author} {\bibinfo {author} {\bibfnamefont {V.}~\bibnamefont
  {Baltz}}, \bibinfo {author} {\bibfnamefont {A.}~\bibnamefont {Manchon}},
  \bibinfo {author} {\bibfnamefont {M.}~\bibnamefont {Tsoi}}, \bibinfo {author}
  {\bibfnamefont {T.}~\bibnamefont {Moriyama}}, \bibinfo {author}
  {\bibfnamefont {T.}~\bibnamefont {Ono}},\ and\ \bibinfo {author}
  {\bibfnamefont {Y.}~\bibnamefont {Tserkovnyak}},\ }\href
  {https://doi.org/10.1103/RevModPhys.90.015005} {\bibfield  {journal}
  {\bibinfo  {journal} {Rev. Mod. Phys.}\ }\textbf {\bibinfo {volume} {90}},\
  \bibinfo {pages} {015005} (\bibinfo {year} {2018})}\BibitemShut {NoStop}%
\bibitem [{\citenamefont {Jungwirth}\ \emph {et~al.}(2016)\citenamefont
  {Jungwirth}, \citenamefont {Marti}, \citenamefont {Wadley},\ and\
  \citenamefont {Wunderlich}}]{Jungwirth2016NN}%
  \BibitemOpen
  \bibfield  {author} {\bibinfo {author} {\bibfnamefont {T.}~\bibnamefont
  {Jungwirth}}, \bibinfo {author} {\bibfnamefont {X.}~\bibnamefont {Marti}},
  \bibinfo {author} {\bibfnamefont {P.}~\bibnamefont {Wadley}},\ and\ \bibinfo
  {author} {\bibfnamefont {J.}~\bibnamefont {Wunderlich}},\ }\href
  {https://doi.org/10.1038/nnano.2016.18} {\bibfield  {journal} {\bibinfo
  {journal} {Nature Nanotechnology}\ }\textbf {\bibinfo {volume} {11}},\
  \bibinfo {pages} {231} (\bibinfo {year} {2016})}\BibitemShut {NoStop}%
\bibitem [{\citenamefont {Nagaosa}\ \emph {et~al.}(2010)\citenamefont
  {Nagaosa}, \citenamefont {Sinova}, \citenamefont {Onoda}, \citenamefont
  {MacDonald},\ and\ \citenamefont {Ong}}]{Nagaosa2010RMP}%
  \BibitemOpen
  \bibfield  {author} {\bibinfo {author} {\bibfnamefont {N.}~\bibnamefont
  {Nagaosa}}, \bibinfo {author} {\bibfnamefont {J.}~\bibnamefont {Sinova}},
  \bibinfo {author} {\bibfnamefont {S.}~\bibnamefont {Onoda}}, \bibinfo
  {author} {\bibfnamefont {A.~H.}\ \bibnamefont {MacDonald}},\ and\ \bibinfo
  {author} {\bibfnamefont {N.~P.}\ \bibnamefont {Ong}},\ }\href
  {https://doi.org/10.1103/RevModPhys.82.1539} {\bibfield  {journal} {\bibinfo
  {journal} {Rev. Mod. Phys.}\ }\textbf {\bibinfo {volume} {82}},\ \bibinfo
  {pages} {1539} (\bibinfo {year} {2010})}\BibitemShut {NoStop}%
\bibitem [{\citenamefont {Tokura}\ \emph {et~al.}(2019)\citenamefont {Tokura},
  \citenamefont {Yasuda},\ and\ \citenamefont {Tsukazaki}}]{Tokura2019NRP}%
  \BibitemOpen
  \bibfield  {author} {\bibinfo {author} {\bibfnamefont {Y.}~\bibnamefont
  {Tokura}}, \bibinfo {author} {\bibfnamefont {K.}~\bibnamefont {Yasuda}},\
  and\ \bibinfo {author} {\bibfnamefont {A.}~\bibnamefont {Tsukazaki}},\ }\href
  {https://doi.org/10.1038/s42254-018-0011-5} {\bibfield  {journal} {\bibinfo
  {journal} {Nature Reviews Physics}\ }\textbf {\bibinfo {volume} {1}},\
  \bibinfo {pages} {126} (\bibinfo {year} {2019})}\BibitemShut {NoStop}%
\bibitem [{\citenamefont {Auerbach}(1994)}]{AuerbachMag}%
  \BibitemOpen
  \bibfield  {author} {\bibinfo {author} {\bibfnamefont {A.}~\bibnamefont
  {Auerbach}},\ }\href@noop {} {\emph {\bibinfo {title} {Interacting Electrons
  and Quantum Magnetism}}}\ (\bibinfo  {publisher} {Springer-Verlag,
  Cambridge},\ \bibinfo {year} {1994})\BibitemShut {NoStop}%
\bibitem [{\citenamefont {Provost}\ and\ \citenamefont
  {Vallee}(1980)}]{Provost1980CMP}%
  \BibitemOpen
  \bibfield  {author} {\bibinfo {author} {\bibfnamefont {J.~P.}\ \bibnamefont
  {Provost}}\ and\ \bibinfo {author} {\bibfnamefont {G.}~\bibnamefont
  {Vallee}},\ }\href {https://doi.org/10.1007/BF02193559} {\bibfield  {journal}
  {\bibinfo  {journal} {Communications in Mathematical Physics}\ }\textbf
  {\bibinfo {volume} {76}},\ \bibinfo {pages} {289} (\bibinfo {year}
  {1980})}\BibitemShut {NoStop}%
\bibitem [{\citenamefont {Ma}\ \emph {et~al.}(2010)\citenamefont {Ma},
  \citenamefont {Chen}, \citenamefont {Fan},\ and\ \citenamefont
  {Liu}}]{MaYQ2010PRB}%
  \BibitemOpen
  \bibfield  {author} {\bibinfo {author} {\bibfnamefont {Y.-Q.}\ \bibnamefont
  {Ma}}, \bibinfo {author} {\bibfnamefont {S.}~\bibnamefont {Chen}}, \bibinfo
  {author} {\bibfnamefont {H.}~\bibnamefont {Fan}},\ and\ \bibinfo {author}
  {\bibfnamefont {W.-M.}\ \bibnamefont {Liu}},\ }\href
  {https://doi.org/10.1103/PhysRevB.81.245129} {\bibfield  {journal} {\bibinfo
  {journal} {Phys. Rev. B}\ }\textbf {\bibinfo {volume} {81}},\ \bibinfo
  {pages} {245129} (\bibinfo {year} {2010})}\BibitemShut {NoStop}%
\bibitem [{\citenamefont {Avron}\ \emph {et~al.}(1995)\citenamefont {Avron},
  \citenamefont {Seiler},\ and\ \citenamefont {Zograf}}]{Avron1995PRL}%
  \BibitemOpen
  \bibfield  {author} {\bibinfo {author} {\bibfnamefont {J.~E.}\ \bibnamefont
  {Avron}}, \bibinfo {author} {\bibfnamefont {R.}~\bibnamefont {Seiler}},\ and\
  \bibinfo {author} {\bibfnamefont {P.~G.}\ \bibnamefont {Zograf}},\ }\href
  {https://doi.org/10.1103/PhysRevLett.75.697} {\bibfield  {journal} {\bibinfo
  {journal} {Phys. Rev. Lett.}\ }\textbf {\bibinfo {volume} {75}},\ \bibinfo
  {pages} {697} (\bibinfo {year} {1995})}\BibitemShut {NoStop}%
\bibitem [{\citenamefont {Xiao}\ \emph {et~al.}(2010)\citenamefont {Xiao},
  \citenamefont {Chang},\ and\ \citenamefont {Niu}}]{XiaoD2010RMP}%
  \BibitemOpen
  \bibfield  {author} {\bibinfo {author} {\bibfnamefont {D.}~\bibnamefont
  {Xiao}}, \bibinfo {author} {\bibfnamefont {M.-C.}\ \bibnamefont {Chang}},\
  and\ \bibinfo {author} {\bibfnamefont {Q.}~\bibnamefont {Niu}},\ }\href
  {https://doi.org/10.1103/RevModPhys.82.1959} {\bibfield  {journal} {\bibinfo
  {journal} {Rev. Mod. Phys.}\ }\textbf {\bibinfo {volume} {82}},\ \bibinfo
  {pages} {1959} (\bibinfo {year} {2010})}\BibitemShut {NoStop}%
\bibitem [{\citenamefont {Tokatly}\ and\ \citenamefont
  {Vignale}(2007)}]{Tokatly2007PRB}%
  \BibitemOpen
  \bibfield  {author} {\bibinfo {author} {\bibfnamefont {I.~V.}\ \bibnamefont
  {Tokatly}}\ and\ \bibinfo {author} {\bibfnamefont {G.}~\bibnamefont
  {Vignale}},\ }\href {https://doi.org/10.1103/PhysRevB.76.161305} {\bibfield
  {journal} {\bibinfo  {journal} {Phys. Rev. B}\ }\textbf {\bibinfo {volume}
  {76}},\ \bibinfo {pages} {161305} (\bibinfo {year} {2007})}\BibitemShut
  {NoStop}%
\bibitem [{\citenamefont {Haldane}(2011)}]{Haldane2011PRL}%
  \BibitemOpen
  \bibfield  {author} {\bibinfo {author} {\bibfnamefont {F.~D.~M.}\
  \bibnamefont {Haldane}},\ }\href
  {https://doi.org/10.1103/PhysRevLett.107.116801} {\bibfield  {journal}
  {\bibinfo  {journal} {Phys. Rev. Lett.}\ }\textbf {\bibinfo {volume} {107}},\
  \bibinfo {pages} {116801} (\bibinfo {year} {2011})}\BibitemShut {NoStop}%
\bibitem [{\citenamefont {Read}\ and\ \citenamefont
  {Rezayi}(2011)}]{Read2011PRB}%
  \BibitemOpen
  \bibfield  {author} {\bibinfo {author} {\bibfnamefont {N.}~\bibnamefont
  {Read}}\ and\ \bibinfo {author} {\bibfnamefont {E.~H.}\ \bibnamefont
  {Rezayi}},\ }\href {https://doi.org/10.1103/PhysRevB.84.085316} {\bibfield
  {journal} {\bibinfo  {journal} {Phys. Rev. B}\ }\textbf {\bibinfo {volume}
  {84}},\ \bibinfo {pages} {085316} (\bibinfo {year} {2011})}\BibitemShut
  {NoStop}%
\bibitem [{\citenamefont {Hughes}\ \emph {et~al.}(2011)\citenamefont {Hughes},
  \citenamefont {Leigh},\ and\ \citenamefont {Fradkin}}]{Hughes2011PRL}%
  \BibitemOpen
  \bibfield  {author} {\bibinfo {author} {\bibfnamefont {T.~L.}\ \bibnamefont
  {Hughes}}, \bibinfo {author} {\bibfnamefont {R.~G.}\ \bibnamefont {Leigh}},\
  and\ \bibinfo {author} {\bibfnamefont {E.}~\bibnamefont {Fradkin}},\ }\href
  {https://doi.org/10.1103/PhysRevLett.107.075502} {\bibfield  {journal}
  {\bibinfo  {journal} {Phys. Rev. Lett.}\ }\textbf {\bibinfo {volume} {107}},\
  \bibinfo {pages} {075502} (\bibinfo {year} {2011})}\BibitemShut {NoStop}%
\bibitem [{\citenamefont {Shapourian}\ \emph {et~al.}(2015)\citenamefont
  {Shapourian}, \citenamefont {Hughes},\ and\ \citenamefont
  {Ryu}}]{Shapourian2015PRB}%
  \BibitemOpen
  \bibfield  {author} {\bibinfo {author} {\bibfnamefont {H.}~\bibnamefont
  {Shapourian}}, \bibinfo {author} {\bibfnamefont {T.~L.}\ \bibnamefont
  {Hughes}},\ and\ \bibinfo {author} {\bibfnamefont {S.}~\bibnamefont {Ryu}},\
  }\href {https://doi.org/10.1103/PhysRevB.92.165131} {\bibfield  {journal}
  {\bibinfo  {journal} {Phys. Rev. B}\ }\textbf {\bibinfo {volume} {92}},\
  \bibinfo {pages} {165131} (\bibinfo {year} {2015})}\BibitemShut {NoStop}%
\bibitem [{\citenamefont {Rao}\ and\ \citenamefont
  {Bradlyn}(2020)}]{Rao2020PRX}%
  \BibitemOpen
  \bibfield  {author} {\bibinfo {author} {\bibfnamefont {P.}~\bibnamefont
  {Rao}}\ and\ \bibinfo {author} {\bibfnamefont {B.}~\bibnamefont {Bradlyn}},\
  }\href {https://doi.org/10.1103/PhysRevX.10.021005} {\bibfield  {journal}
  {\bibinfo  {journal} {Phys. Rev. X}\ }\textbf {\bibinfo {volume} {10}},\
  \bibinfo {pages} {021005} (\bibinfo {year} {2020})}\BibitemShut {NoStop}%
\bibitem [{\citenamefont {Hoyos}\ and\ \citenamefont
  {Son}(2012)}]{Hoyos2012PRL}%
  \BibitemOpen
  \bibfield  {author} {\bibinfo {author} {\bibfnamefont {C.}~\bibnamefont
  {Hoyos}}\ and\ \bibinfo {author} {\bibfnamefont {D.~T.}\ \bibnamefont
  {Son}},\ }\href {https://doi.org/10.1103/PhysRevLett.108.066805} {\bibfield
  {journal} {\bibinfo  {journal} {Phys. Rev. Lett.}\ }\textbf {\bibinfo
  {volume} {108}},\ \bibinfo {pages} {066805} (\bibinfo {year}
  {2012})}\BibitemShut {NoStop}%
\bibitem [{\citenamefont {Kozii}\ \emph {et~al.}(2021)\citenamefont {Kozii},
  \citenamefont {Avdoshkin}, \citenamefont {Zhong},\ and\ \citenamefont
  {Moore}}]{Kozii2021PRL}%
  \BibitemOpen
  \bibfield  {author} {\bibinfo {author} {\bibfnamefont {V.}~\bibnamefont
  {Kozii}}, \bibinfo {author} {\bibfnamefont {A.}~\bibnamefont {Avdoshkin}},
  \bibinfo {author} {\bibfnamefont {S.}~\bibnamefont {Zhong}},\ and\ \bibinfo
  {author} {\bibfnamefont {J.~E.}\ \bibnamefont {Moore}},\ }\href
  {https://doi.org/10.1103/PhysRevLett.126.156602} {\bibfield  {journal}
  {\bibinfo  {journal} {Phys. Rev. Lett.}\ }\textbf {\bibinfo {volume} {126}},\
  \bibinfo {pages} {156602} (\bibinfo {year} {2021})}\BibitemShut {NoStop}%
\bibitem [{\citenamefont {Barkeshli}\ \emph {et~al.}(2012)\citenamefont
  {Barkeshli}, \citenamefont {Chung},\ and\ \citenamefont
  {Qi}}]{Barkeshli2012PRB}%
  \BibitemOpen
  \bibfield  {author} {\bibinfo {author} {\bibfnamefont {M.}~\bibnamefont
  {Barkeshli}}, \bibinfo {author} {\bibfnamefont {S.~B.}\ \bibnamefont
  {Chung}},\ and\ \bibinfo {author} {\bibfnamefont {X.-L.}\ \bibnamefont
  {Qi}},\ }\href {https://doi.org/10.1103/PhysRevB.85.245107} {\bibfield
  {journal} {\bibinfo  {journal} {Phys. Rev. B}\ }\textbf {\bibinfo {volume}
  {85}},\ \bibinfo {pages} {245107} (\bibinfo {year} {2012})}\BibitemShut
  {NoStop}%
\bibitem [{\citenamefont {Liu}\ and\ \citenamefont {Shi}(2017)}]{LiuDH2017PRL}%
  \BibitemOpen
  \bibfield  {author} {\bibinfo {author} {\bibfnamefont {D.}~\bibnamefont
  {Liu}}\ and\ \bibinfo {author} {\bibfnamefont {J.}~\bibnamefont {Shi}},\
  }\href {https://doi.org/10.1103/PhysRevLett.119.075301} {\bibfield  {journal}
  {\bibinfo  {journal} {Phys. Rev. Lett.}\ }\textbf {\bibinfo {volume} {119}},\
  \bibinfo {pages} {075301} (\bibinfo {year} {2017})}\BibitemShut {NoStop}%
\bibitem [{\citenamefont {{Li}}\ \emph {et~al.}(2025)\citenamefont {{Li}},
  \citenamefont {{Yang}}, \citenamefont {{Qin}}, \citenamefont {{Zhou}},\ and\
  \citenamefont {{Yao}}}]{LiD2025arXiv}%
  \BibitemOpen
  \bibfield  {author} {\bibinfo {author} {\bibfnamefont {D.}~\bibnamefont
  {{Li}}}, \bibinfo {author} {\bibfnamefont {G.}~\bibnamefont {{Yang}}},
  \bibinfo {author} {\bibfnamefont {T.}~\bibnamefont {{Qin}}}, \bibinfo
  {author} {\bibfnamefont {J.}~\bibnamefont {{Zhou}}},\ and\ \bibinfo {author}
  {\bibfnamefont {Y.}~\bibnamefont {{Yao}}},\ }\href
  {https://arxiv.org/abs/2601.13283} {\bibfield  {journal} {\bibinfo  {journal}
  {arXiv: 2511.16141}\ } (\bibinfo {year} {2025})}\BibitemShut {NoStop}%
\bibitem [{EPh()}]{EPhC}%
  \BibitemOpen
  \href@noop {} {\bibinfo  {journal} {we choose the electron-phonon coupling,
  which can be cast into a vector coupling form as
  $\hat{H}_{e-ph}=g\hat{\bm{v}}\cdot\hat{\bm{A}}\left[\bm{u}\right]$, where $g$
  is the coupling constant, $\hat{\bm{v}}$ is the electron velocity, and
  $\hat{\bm{A}}\left[\bm{u}\right]$ is the vector potential due to strain
  $\bm{u}$~\citep{Mead1979JCP,QinT2012PRB,HuLH2021PRL,ShanWY2022PRB,HuJM2025PRL}}\
  }\BibitemShut {NoStop}%
\bibitem [{\citenamefont {Nakahara}(2003)}]{Nakahara2003GTP}%
  \BibitemOpen
\bibfield  {journal} {  }\bibfield  {author} {\bibinfo {author} {\bibfnamefont
  {M.}~\bibnamefont {Nakahara}},\ }\href
  {https://doi.org/10.1201/9781315275826} {\emph {\bibinfo {title} {Geometry,
  Topology and Physics}}},\ \bibinfo {edition} {2nd}\ ed.\ (\bibinfo
  {publisher} {CRC Press},\ \bibinfo {address} {Boca Raton},\ \bibinfo {year}
  {2003})\BibitemShut {NoStop}%
\bibitem [{\citenamefont {Zhang}\ and\ \citenamefont
  {Niu}(2015)}]{ZhangLF2015PRL}%
  \BibitemOpen
  \bibfield  {author} {\bibinfo {author} {\bibfnamefont {L.}~\bibnamefont
  {Zhang}}\ and\ \bibinfo {author} {\bibfnamefont {Q.}~\bibnamefont {Niu}},\
  }\href {https://doi.org/10.1103/PhysRevLett.115.115502} {\bibfield  {journal}
  {\bibinfo  {journal} {Phys. Rev. Lett.}\ }\textbf {\bibinfo {volume} {115}},\
  \bibinfo {pages} {115502} (\bibinfo {year} {2015})}\BibitemShut {NoStop}%
\bibitem [{\citenamefont {Saparov}\ \emph {et~al.}(2022)\citenamefont
  {Saparov}, \citenamefont {Xiong}, \citenamefont {Ren},\ and\ \citenamefont
  {Niu}}]{Saparov2022PRB}%
  \BibitemOpen
  \bibfield  {author} {\bibinfo {author} {\bibfnamefont {D.}~\bibnamefont
  {Saparov}}, \bibinfo {author} {\bibfnamefont {B.}~\bibnamefont {Xiong}},
  \bibinfo {author} {\bibfnamefont {Y.}~\bibnamefont {Ren}},\ and\ \bibinfo
  {author} {\bibfnamefont {Q.}~\bibnamefont {Niu}},\ }\href
  {https://doi.org/10.1103/PhysRevB.105.064303} {\bibfield  {journal} {\bibinfo
   {journal} {Phys. Rev. B}\ }\textbf {\bibinfo {volume} {105}},\ \bibinfo
  {pages} {064303} (\bibinfo {year} {2022})}\BibitemShut {NoStop}%
\bibitem [{\citenamefont {Zhang}\ and\ \citenamefont
  {Murakami}(2022)}]{ZhangTT2022PRR}%
  \BibitemOpen
  \bibfield  {author} {\bibinfo {author} {\bibfnamefont {T.}~\bibnamefont
  {Zhang}}\ and\ \bibinfo {author} {\bibfnamefont {S.}~\bibnamefont
  {Murakami}},\ }\href {https://doi.org/10.1103/PhysRevResearch.4.L012024}
  {\bibfield  {journal} {\bibinfo  {journal} {Phys. Rev. Res.}\ }\textbf
  {\bibinfo {volume} {4}},\ \bibinfo {pages} {L012024} (\bibinfo {year}
  {2022})}\BibitemShut {NoStop}%
\bibitem [{\citenamefont {Chen}\ \emph {et~al.}(2025)\citenamefont {Chen},
  \citenamefont {Qin}, \citenamefont {Zhang}, \citenamefont {Cui},
  \citenamefont {Niu},\ and\ \citenamefont {Zhang}}]{ChenYR2025PRL}%
  \BibitemOpen
  \bibfield  {author} {\bibinfo {author} {\bibfnamefont {Y.}~\bibnamefont
  {Chen}}, \bibinfo {author} {\bibfnamefont {W.}~\bibnamefont {Qin}}, \bibinfo
  {author} {\bibfnamefont {S.}~\bibnamefont {Zhang}}, \bibinfo {author}
  {\bibfnamefont {P.}~\bibnamefont {Cui}}, \bibinfo {author} {\bibfnamefont
  {Q.}~\bibnamefont {Niu}},\ and\ \bibinfo {author} {\bibfnamefont
  {Z.}~\bibnamefont {Zhang}},\ }\href {https://doi.org/10.1103/bfll-sdrb}
  {\bibfield  {journal} {\bibinfo  {journal} {Phys. Rev. Lett.}\ }\textbf
  {\bibinfo {volume} {135}},\ \bibinfo {pages} {126608} (\bibinfo {year}
  {2025})}\BibitemShut {NoStop}%
\bibitem [{\citenamefont {Chatterjee}\ and\ \citenamefont
  {Liu}(2026)}]{chatterjee2026arxiv}%
  \BibitemOpen
  \bibfield  {author} {\bibinfo {author} {\bibfnamefont {A.}~\bibnamefont
  {Chatterjee}}\ and\ \bibinfo {author} {\bibfnamefont {C.-X.}\ \bibnamefont
  {Liu}},\ }\href {https://arxiv.org/abs/2601.13283} {\bibfield  {journal}
  {\bibinfo  {journal} {arXiv: 2601.13283}\ } (\bibinfo {year}
  {2026})}\BibitemShut {NoStop}%
\bibitem [{\citenamefont {Chen}\ \emph
  {et~al.}(2026{\natexlab{a}})\citenamefont {Chen}, \citenamefont {Chen},
  \citenamefont {Yang}, \citenamefont {Cao},\ and\ \citenamefont
  {Xiao}}]{ChenHR2026arxiv}%
  \BibitemOpen
  \bibfield  {author} {\bibinfo {author} {\bibfnamefont {H.}~\bibnamefont
  {Chen}}, \bibinfo {author} {\bibfnamefont {W.}~\bibnamefont {Chen}}, \bibinfo
  {author} {\bibfnamefont {K.}~\bibnamefont {Yang}}, \bibinfo {author}
  {\bibfnamefont {T.}~\bibnamefont {Cao}},\ and\ \bibinfo {author}
  {\bibfnamefont {D.}~\bibnamefont {Xiao}},\ }\href
  {https://arxiv.org/abs/2605.06983} {\bibfield  {journal} {\bibinfo  {journal}
  {arXiv: 2605.06983}\ } (\bibinfo {year} {2026}{\natexlab{a}})}\BibitemShut
  {NoStop}%
\bibitem [{\citenamefont {Onishi}\ and\ \citenamefont
  {Fu}(2024)}]{Onishi2024PRX}%
  \BibitemOpen
  \bibfield  {author} {\bibinfo {author} {\bibfnamefont {Y.}~\bibnamefont
  {Onishi}}\ and\ \bibinfo {author} {\bibfnamefont {L.}~\bibnamefont {Fu}},\
  }\href {https://doi.org/10.1103/PhysRevX.14.011052} {\bibfield  {journal}
  {\bibinfo  {journal} {Phys. Rev. X}\ }\textbf {\bibinfo {volume} {14}},\
  \bibinfo {pages} {011052} (\bibinfo {year} {2024})}\BibitemShut {NoStop}%
\bibitem [{\citenamefont {Pai}\ and\ \citenamefont {Zhang}(2026)}]{Pai2026PRL}%
  \BibitemOpen
  \bibfield  {author} {\bibinfo {author} {\bibfnamefont {P.}~\bibnamefont
  {Pai}}\ and\ \bibinfo {author} {\bibfnamefont {F.}~\bibnamefont {Zhang}},\
  }\href {https://doi.org/10.1103/h9vk-lrvk} {\bibfield  {journal} {\bibinfo
  {journal} {Phys. Rev. Lett.}\ }\textbf {\bibinfo {volume} {136}},\ \bibinfo
  {pages} {116601} (\bibinfo {year} {2026})}\BibitemShut {NoStop}%
\bibitem [{CSI()}]{CSIneq}%
  \BibitemOpen
  \href@noop {} {\bibinfo  {journal} {The Cauchy-Schwarz inequality for the
  bounded and summable functions $f_{1}\left(x\right)$ and
  $f_{2}\left(x\right)$ has the form $\left(\int
  f_{1}\left(x\right)f_{2}\left(x\right)dx\right)^{2}\leq\left(\int
  f_{1}^{2}\left(x\right)dx\right)\times\left(\int
  f_{2}^{2}\left(x\right)dx\right)$~\citep{Riesz1955FA}}\ }\BibitemShut
  {NoStop}%
\bibitem [{\citenamefont {Roy}(2014)}]{Roy2014PRB}%
  \BibitemOpen
\bibfield  {journal} {  }\bibfield  {author} {\bibinfo {author} {\bibfnamefont
  {R.}~\bibnamefont {Roy}},\ }\href
  {https://doi.org/10.1103/PhysRevB.90.165139} {\bibfield  {journal} {\bibinfo
  {journal} {Phys. Rev. B}\ }\textbf {\bibinfo {volume} {90}},\ \bibinfo
  {pages} {165139} (\bibinfo {year} {2014})}\BibitemShut {NoStop}%
\bibitem [{\citenamefont {Ozawa}\ and\ \citenamefont
  {Mera}(2021)}]{Ozawa2021PRB}%
  \BibitemOpen
  \bibfield  {author} {\bibinfo {author} {\bibfnamefont {T.}~\bibnamefont
  {Ozawa}}\ and\ \bibinfo {author} {\bibfnamefont {B.}~\bibnamefont {Mera}},\
  }\href {https://doi.org/10.1103/PhysRevB.104.045103} {\bibfield  {journal}
  {\bibinfo  {journal} {Phys. Rev. B}\ }\textbf {\bibinfo {volume} {104}},\
  \bibinfo {pages} {045103} (\bibinfo {year} {2021})}\BibitemShut {NoStop}%
\bibitem [{\citenamefont {Julku}\ \emph {et~al.}(2016)\citenamefont {Julku},
  \citenamefont {Peotta}, \citenamefont {Vanhala}, \citenamefont {Kim},\ and\
  \citenamefont {T\"orm\"a}}]{Julku2016PRL}%
  \BibitemOpen
  \bibfield  {author} {\bibinfo {author} {\bibfnamefont {A.}~\bibnamefont
  {Julku}}, \bibinfo {author} {\bibfnamefont {S.}~\bibnamefont {Peotta}},
  \bibinfo {author} {\bibfnamefont {T.~I.}\ \bibnamefont {Vanhala}}, \bibinfo
  {author} {\bibfnamefont {D.-H.}\ \bibnamefont {Kim}},\ and\ \bibinfo {author}
  {\bibfnamefont {P.}~\bibnamefont {T\"orm\"a}},\ }\href
  {https://doi.org/10.1103/PhysRevLett.117.045303} {\bibfield  {journal}
  {\bibinfo  {journal} {Phys. Rev. Lett.}\ }\textbf {\bibinfo {volume} {117}},\
  \bibinfo {pages} {045303} (\bibinfo {year} {2016})}\BibitemShut {NoStop}%
\bibitem [{\citenamefont {Santini}\ \emph {et~al.}(2009)\citenamefont
  {Santini}, \citenamefont {Carretta}, \citenamefont {Amoretti}, \citenamefont
  {Caciuffo}, \citenamefont {Magnani},\ and\ \citenamefont
  {Lander}}]{Santini2009RMP}%
  \BibitemOpen
  \bibfield  {author} {\bibinfo {author} {\bibfnamefont {P.}~\bibnamefont
  {Santini}}, \bibinfo {author} {\bibfnamefont {S.}~\bibnamefont {Carretta}},
  \bibinfo {author} {\bibfnamefont {G.}~\bibnamefont {Amoretti}}, \bibinfo
  {author} {\bibfnamefont {R.}~\bibnamefont {Caciuffo}}, \bibinfo {author}
  {\bibfnamefont {N.}~\bibnamefont {Magnani}},\ and\ \bibinfo {author}
  {\bibfnamefont {G.~H.}\ \bibnamefont {Lander}},\ }\href
  {https://doi.org/10.1103/RevModPhys.81.807} {\bibfield  {journal} {\bibinfo
  {journal} {Rev. Mod. Phys.}\ }\textbf {\bibinfo {volume} {81}},\ \bibinfo
  {pages} {807} (\bibinfo {year} {2009})}\BibitemShut {NoStop}%
\bibitem [{\citenamefont {Sakurai}\ and\ \citenamefont
  {Napolitano}(2020)}]{Sakurai2020MQM}%
  \BibitemOpen
  \bibfield  {author} {\bibinfo {author} {\bibfnamefont {J.~J.}\ \bibnamefont
  {Sakurai}}\ and\ \bibinfo {author} {\bibfnamefont {J.}~\bibnamefont
  {Napolitano}},\ }\href@noop {} {\emph {\bibinfo {title} {Modern Quantum
  Mechanics}}},\ \bibinfo {edition} {3rd}\ ed.\ (\bibinfo  {publisher}
  {Cambridge University Press},\ \bibinfo {year} {2020})\BibitemShut {NoStop}%
\bibitem [{\citenamefont {Tahir}\ and\ \citenamefont
  {Chen}(2023)}]{Tahir2023PRL}%
  \BibitemOpen
  \bibfield  {author} {\bibinfo {author} {\bibfnamefont {M.}~\bibnamefont
  {Tahir}}\ and\ \bibinfo {author} {\bibfnamefont {H.}~\bibnamefont {Chen}},\
  }\href {https://doi.org/10.1103/PhysRevLett.131.106701} {\bibfield  {journal}
  {\bibinfo  {journal} {Phys. Rev. Lett.}\ }\textbf {\bibinfo {volume} {131}},\
  \bibinfo {pages} {106701} (\bibinfo {year} {2023})}\BibitemShut {NoStop}%
\bibitem [{SMs()}]{SMs}%
  \BibitemOpen
  \href@noop {} {\bibinfo  {journal} {See Supplemental Material for details of
  multipole expansion of Berry curvature, symmetry properties of EHV and AHC
  for all 10 MLGs and model calculations}\ }\BibitemShut {NoStop}%
\bibitem [{\citenamefont {Suzuki}\ \emph {et~al.}(2017)\citenamefont {Suzuki},
  \citenamefont {Koretsune}, \citenamefont {Ochi},\ and\ \citenamefont
  {Arita}}]{Suzuki2017PRB}%
  \BibitemOpen
\bibfield  {journal} {  }\bibfield  {author} {\bibinfo {author} {\bibfnamefont
  {M.-T.}\ \bibnamefont {Suzuki}}, \bibinfo {author} {\bibfnamefont
  {T.}~\bibnamefont {Koretsune}}, \bibinfo {author} {\bibfnamefont
  {M.}~\bibnamefont {Ochi}},\ and\ \bibinfo {author} {\bibfnamefont
  {R.}~\bibnamefont {Arita}},\ }\href
  {https://doi.org/10.1103/PhysRevB.95.094406} {\bibfield  {journal} {\bibinfo
  {journal} {Phys. Rev. B}\ }\textbf {\bibinfo {volume} {95}},\ \bibinfo
  {pages} {094406} (\bibinfo {year} {2017})}\BibitemShut {NoStop}%
\bibitem [{\citenamefont {Cao}\ \emph {et~al.}(2023)\citenamefont {Cao},
  \citenamefont {Jiang}, \citenamefont {Li}, \citenamefont {Tu}, \citenamefont
  {Zhou}, \citenamefont {Zhou},\ and\ \citenamefont {Yao}}]{CaoJ2023PRL}%
  \BibitemOpen
  \bibfield  {author} {\bibinfo {author} {\bibfnamefont {J.}~\bibnamefont
  {Cao}}, \bibinfo {author} {\bibfnamefont {W.}~\bibnamefont {Jiang}}, \bibinfo
  {author} {\bibfnamefont {X.-P.}\ \bibnamefont {Li}}, \bibinfo {author}
  {\bibfnamefont {D.}~\bibnamefont {Tu}}, \bibinfo {author} {\bibfnamefont
  {J.}~\bibnamefont {Zhou}}, \bibinfo {author} {\bibfnamefont {J.}~\bibnamefont
  {Zhou}},\ and\ \bibinfo {author} {\bibfnamefont {Y.}~\bibnamefont {Yao}},\
  }\href {https://doi.org/10.1103/PhysRevLett.130.166702} {\bibfield  {journal}
  {\bibinfo  {journal} {Phys. Rev. Lett.}\ }\textbf {\bibinfo {volume} {130}},\
  \bibinfo {pages} {166702} (\bibinfo {year} {2023})}\BibitemShut {NoStop}%
\bibitem [{\citenamefont {Kurumaji}(2023)}]{Kurumaji2023PRR}%
  \BibitemOpen
  \bibfield  {author} {\bibinfo {author} {\bibfnamefont {T.}~\bibnamefont
  {Kurumaji}},\ }\href {https://doi.org/10.1103/PhysRevResearch.5.023138}
  {\bibfield  {journal} {\bibinfo  {journal} {Phys. Rev. Res.}\ }\textbf
  {\bibinfo {volume} {5}},\ \bibinfo {pages} {023138} (\bibinfo {year}
  {2023})}\BibitemShut {NoStop}%
\bibitem [{\citenamefont {Kleiner}(1966)}]{Kleiner1966PR}%
  \BibitemOpen
  \bibfield  {author} {\bibinfo {author} {\bibfnamefont {W.~H.}\ \bibnamefont
  {Kleiner}},\ }\href {https://doi.org/10.1103/PhysRev.142.318} {\bibfield
  {journal} {\bibinfo  {journal} {Phys. Rev.}\ }\textbf {\bibinfo {volume}
  {142}},\ \bibinfo {pages} {318} (\bibinfo {year} {1966})}\BibitemShut
  {NoStop}%
\bibitem [{\citenamefont {Zhou}\ \emph {et~al.}(2022)\citenamefont {Zhou},
  \citenamefont {Zhang}, \citenamefont {Lin}, \citenamefont {Cao},
  \citenamefont {Zhou}, \citenamefont {Jiang}, \citenamefont {Du},
  \citenamefont {Tang}, \citenamefont {Shi}, \citenamefont {Jiang},
  \citenamefont {Cao}, \citenamefont {Lin}, \citenamefont {Fu}, \citenamefont
  {Zhu}, \citenamefont {Guo}, \citenamefont {Huang}, \citenamefont {Yao},
  \citenamefont {Parkin}, \citenamefont {Zhou}, \citenamefont {Gao},
  \citenamefont {Wang}, \citenamefont {Hou}, \citenamefont {Yao}, \citenamefont
  {Suenaga}, \citenamefont {Wu},\ and\ \citenamefont {Liu}}]{ZhouJD2022Nature}%
  \BibitemOpen
  \bibfield  {author} {\bibinfo {author} {\bibfnamefont {J.}~\bibnamefont
  {Zhou}}, \bibinfo {author} {\bibfnamefont {W.}~\bibnamefont {Zhang}},
  \bibinfo {author} {\bibfnamefont {Y.-C.}\ \bibnamefont {Lin}}, \bibinfo
  {author} {\bibfnamefont {J.}~\bibnamefont {Cao}}, \bibinfo {author}
  {\bibfnamefont {Y.}~\bibnamefont {Zhou}}, \bibinfo {author} {\bibfnamefont
  {W.}~\bibnamefont {Jiang}}, \bibinfo {author} {\bibfnamefont
  {H.}~\bibnamefont {Du}}, \bibinfo {author} {\bibfnamefont {B.}~\bibnamefont
  {Tang}}, \bibinfo {author} {\bibfnamefont {J.}~\bibnamefont {Shi}}, \bibinfo
  {author} {\bibfnamefont {B.}~\bibnamefont {Jiang}}, \bibinfo {author}
  {\bibfnamefont {X.}~\bibnamefont {Cao}}, \bibinfo {author} {\bibfnamefont
  {B.}~\bibnamefont {Lin}}, \bibinfo {author} {\bibfnamefont {Q.}~\bibnamefont
  {Fu}}, \bibinfo {author} {\bibfnamefont {C.}~\bibnamefont {Zhu}}, \bibinfo
  {author} {\bibfnamefont {W.}~\bibnamefont {Guo}}, \bibinfo {author}
  {\bibfnamefont {Y.}~\bibnamefont {Huang}}, \bibinfo {author} {\bibfnamefont
  {Y.}~\bibnamefont {Yao}}, \bibinfo {author} {\bibfnamefont {S.~S.~P.}\
  \bibnamefont {Parkin}}, \bibinfo {author} {\bibfnamefont {J.}~\bibnamefont
  {Zhou}}, \bibinfo {author} {\bibfnamefont {Y.}~\bibnamefont {Gao}}, \bibinfo
  {author} {\bibfnamefont {Y.}~\bibnamefont {Wang}}, \bibinfo {author}
  {\bibfnamefont {Y.}~\bibnamefont {Hou}}, \bibinfo {author} {\bibfnamefont
  {Y.}~\bibnamefont {Yao}}, \bibinfo {author} {\bibfnamefont {K.}~\bibnamefont
  {Suenaga}}, \bibinfo {author} {\bibfnamefont {X.}~\bibnamefont {Wu}},\ and\
  \bibinfo {author} {\bibfnamefont {Z.}~\bibnamefont {Liu}},\ }\href
  {https://doi.org/10.1038/s41586-022-05031-2} {\bibfield  {journal} {\bibinfo
  {journal} {Nature}\ }\textbf {\bibinfo {volume} {609}},\ \bibinfo {pages}
  {46} (\bibinfo {year} {2022})}\BibitemShut {NoStop}%
\bibitem [{\citenamefont {Seemann}\ \emph {et~al.}(2015)\citenamefont
  {Seemann}, \citenamefont {K\"odderitzsch}, \citenamefont {Wimmer},\ and\
  \citenamefont {Ebert}}]{Seemann2015PRB}%
  \BibitemOpen
  \bibfield  {author} {\bibinfo {author} {\bibfnamefont {M.}~\bibnamefont
  {Seemann}}, \bibinfo {author} {\bibfnamefont {D.}~\bibnamefont
  {K\"odderitzsch}}, \bibinfo {author} {\bibfnamefont {S.}~\bibnamefont
  {Wimmer}},\ and\ \bibinfo {author} {\bibfnamefont {H.}~\bibnamefont
  {Ebert}},\ }\href {https://doi.org/10.1103/PhysRevB.92.155138} {\bibfield
  {journal} {\bibinfo  {journal} {Phys. Rev. B}\ }\textbf {\bibinfo {volume}
  {92}},\ \bibinfo {pages} {155138} (\bibinfo {year} {2015})}\BibitemShut
  {NoStop}%
\bibitem [{\citenamefont {Sandratskii}\ \emph {et~al.}(1981)\citenamefont
  {Sandratskii}, \citenamefont {Egorov},\ and\ \citenamefont
  {Berdyshev}}]{Sandratskii1981PSSb}%
  \BibitemOpen
  \bibfield  {author} {\bibinfo {author} {\bibfnamefont {L.~M.}\ \bibnamefont
  {Sandratskii}}, \bibinfo {author} {\bibfnamefont {R.~F.}\ \bibnamefont
  {Egorov}},\ and\ \bibinfo {author} {\bibfnamefont {A.~A.}\ \bibnamefont
  {Berdyshev}},\ }\href
  {https://doi.org/https://doi.org/10.1002/pssb.2221040111} {\bibfield
  {journal} {\bibinfo  {journal} {physica status solidi (b)}\ }\textbf
  {\bibinfo {volume} {104}},\ \bibinfo {pages} {103} (\bibinfo {year}
  {1981})}\BibitemShut {NoStop}%
\bibitem [{\citenamefont {Wu}\ \emph {et~al.}(2007)\citenamefont {Wu},
  \citenamefont {Sun}, \citenamefont {Fradkin},\ and\ \citenamefont
  {Zhang}}]{WuCJ2007PRB}%
  \BibitemOpen
  \bibfield  {author} {\bibinfo {author} {\bibfnamefont {C.}~\bibnamefont
  {Wu}}, \bibinfo {author} {\bibfnamefont {K.}~\bibnamefont {Sun}}, \bibinfo
  {author} {\bibfnamefont {E.}~\bibnamefont {Fradkin}},\ and\ \bibinfo {author}
  {\bibfnamefont {S.-C.}\ \bibnamefont {Zhang}},\ }\href
  {https://doi.org/10.1103/PhysRevB.75.115103} {\bibfield  {journal} {\bibinfo
  {journal} {Phys. Rev. B}\ }\textbf {\bibinfo {volume} {75}},\ \bibinfo
  {pages} {115103} (\bibinfo {year} {2007})}\BibitemShut {NoStop}%
\bibitem [{\citenamefont {Yuan}\ \emph {et~al.}(2020)\citenamefont {Yuan},
  \citenamefont {Wang}, \citenamefont {Luo}, \citenamefont {Rashba},\ and\
  \citenamefont {Zunger}}]{YuanLD2020PRB}%
  \BibitemOpen
  \bibfield  {author} {\bibinfo {author} {\bibfnamefont {L.-D.}\ \bibnamefont
  {Yuan}}, \bibinfo {author} {\bibfnamefont {Z.}~\bibnamefont {Wang}}, \bibinfo
  {author} {\bibfnamefont {J.-W.}\ \bibnamefont {Luo}}, \bibinfo {author}
  {\bibfnamefont {E.~I.}\ \bibnamefont {Rashba}},\ and\ \bibinfo {author}
  {\bibfnamefont {A.}~\bibnamefont {Zunger}},\ }\href
  {https://doi.org/10.1103/PhysRevB.102.014422} {\bibfield  {journal} {\bibinfo
   {journal} {Phys. Rev. B}\ }\textbf {\bibinfo {volume} {102}},\ \bibinfo
  {pages} {014422} (\bibinfo {year} {2020})}\BibitemShut {NoStop}%
\bibitem [{\citenamefont {Šmejkal}\ \emph {et~al.}(2020)\citenamefont
  {Šmejkal}, \citenamefont {González-Hernández}, \citenamefont {Jungwirth},\
  and\ \citenamefont {Sinova}}]{Smejkal2021SA}%
  \BibitemOpen
  \bibfield  {author} {\bibinfo {author} {\bibfnamefont {L.}~\bibnamefont
  {Šmejkal}}, \bibinfo {author} {\bibfnamefont {R.}~\bibnamefont
  {González-Hernández}}, \bibinfo {author} {\bibfnamefont {T.}~\bibnamefont
  {Jungwirth}},\ and\ \bibinfo {author} {\bibfnamefont {J.}~\bibnamefont
  {Sinova}},\ }\href {https://doi.org/10.1126/sciadv.aaz8809} {\bibfield
  {journal} {\bibinfo  {journal} {Science Advances}\ }\textbf {\bibinfo
  {volume} {6}},\ \bibinfo {pages} {eaaz8809} (\bibinfo {year}
  {2020})}\BibitemShut {NoStop}%
\bibitem [{\citenamefont {Ma}\ \emph {et~al.}(2021)\citenamefont {Ma},
  \citenamefont {Hu}, \citenamefont {Li}, \citenamefont {Liu}, \citenamefont
  {Yao}, \citenamefont {Jia},\ and\ \citenamefont {Liu}}]{MaHY2021NC}%
  \BibitemOpen
  \bibfield  {author} {\bibinfo {author} {\bibfnamefont {H.-Y.}\ \bibnamefont
  {Ma}}, \bibinfo {author} {\bibfnamefont {M.}~\bibnamefont {Hu}}, \bibinfo
  {author} {\bibfnamefont {N.}~\bibnamefont {Li}}, \bibinfo {author}
  {\bibfnamefont {J.}~\bibnamefont {Liu}}, \bibinfo {author} {\bibfnamefont
  {W.}~\bibnamefont {Yao}}, \bibinfo {author} {\bibfnamefont {J.-F.}\
  \bibnamefont {Jia}},\ and\ \bibinfo {author} {\bibfnamefont {J.}~\bibnamefont
  {Liu}},\ }\href {https://doi.org/10.1038/s41467-021-23127-7} {\bibfield
  {journal} {\bibinfo  {journal} {Nature Communications}\ }\textbf {\bibinfo
  {volume} {12}},\ \bibinfo {pages} {2846} (\bibinfo {year}
  {2021})}\BibitemShut {NoStop}%
\bibitem [{\citenamefont {\ifmmode~\check{S}\else \v{S}\fi{}mejkal}\ \emph
  {et~al.}(2022)\citenamefont {\ifmmode~\check{S}\else \v{S}\fi{}mejkal},
  \citenamefont {Sinova},\ and\ \citenamefont {Jungwirth}}]{Smejkal2022PRX3}%
  \BibitemOpen
  \bibfield  {author} {\bibinfo {author} {\bibfnamefont {L.}~\bibnamefont
  {\ifmmode~\check{S}\else \v{S}\fi{}mejkal}}, \bibinfo {author} {\bibfnamefont
  {J.}~\bibnamefont {Sinova}},\ and\ \bibinfo {author} {\bibfnamefont
  {T.}~\bibnamefont {Jungwirth}},\ }\href
  {https://doi.org/10.1103/PhysRevX.12.031042} {\bibfield  {journal} {\bibinfo
  {journal} {Phys. Rev. X}\ }\textbf {\bibinfo {volume} {12}},\ \bibinfo
  {pages} {031042} (\bibinfo {year} {2022})}\BibitemShut {NoStop}%
\bibitem [{\citenamefont {Mazin}(2022)}]{Mazin2022PRX}%
  \BibitemOpen
  \bibfield  {author} {\bibinfo {author} {\bibfnamefont {I.}~\bibnamefont
  {Mazin}},\ }\href {https://doi.org/10.1103/PhysRevX.12.040002} {\bibfield
  {journal} {\bibinfo  {journal} {Phys. Rev. X}\ }\textbf {\bibinfo {volume}
  {12}},\ \bibinfo {pages} {040002} (\bibinfo {year} {2022})}\BibitemShut
  {NoStop}%
\bibitem [{\citenamefont {Bai}\ \emph {et~al.}(2024)\citenamefont {Bai},
  \citenamefont {Feng}, \citenamefont {Liu}, \citenamefont {Šmejkal},
  \citenamefont {Mokrousov},\ and\ \citenamefont {Yao}}]{BaiL2024AFM}%
  \BibitemOpen
  \bibfield  {author} {\bibinfo {author} {\bibfnamefont {L.}~\bibnamefont
  {Bai}}, \bibinfo {author} {\bibfnamefont {W.}~\bibnamefont {Feng}}, \bibinfo
  {author} {\bibfnamefont {S.}~\bibnamefont {Liu}}, \bibinfo {author}
  {\bibfnamefont {L.}~\bibnamefont {Šmejkal}}, \bibinfo {author}
  {\bibfnamefont {Y.}~\bibnamefont {Mokrousov}},\ and\ \bibinfo {author}
  {\bibfnamefont {Y.}~\bibnamefont {Yao}},\ }\href
  {https://doi.org/https://doi.org/10.1002/adfm.202409327} {\bibfield
  {journal} {\bibinfo  {journal} {Advanced Functional Materials}\ }\textbf
  {\bibinfo {volume} {34}},\ \bibinfo {pages} {2409327} (\bibinfo {year}
  {2024})}\BibitemShut {NoStop}%
\bibitem [{\citenamefont {Song}\ \emph {et~al.}(2025)\citenamefont {Song},
  \citenamefont {Bai}, \citenamefont {Zhou}, \citenamefont {Han}, \citenamefont
  {Reichlova}, \citenamefont {Dil}, \citenamefont {Liu}, \citenamefont {Chen},\
  and\ \citenamefont {Pan}}]{SongC2025NRM}%
  \BibitemOpen
  \bibfield  {author} {\bibinfo {author} {\bibfnamefont {C.}~\bibnamefont
  {Song}}, \bibinfo {author} {\bibfnamefont {H.}~\bibnamefont {Bai}}, \bibinfo
  {author} {\bibfnamefont {Z.}~\bibnamefont {Zhou}}, \bibinfo {author}
  {\bibfnamefont {L.}~\bibnamefont {Han}}, \bibinfo {author} {\bibfnamefont
  {H.}~\bibnamefont {Reichlova}}, \bibinfo {author} {\bibfnamefont {J.~H.}\
  \bibnamefont {Dil}}, \bibinfo {author} {\bibfnamefont {J.}~\bibnamefont
  {Liu}}, \bibinfo {author} {\bibfnamefont {X.}~\bibnamefont {Chen}},\ and\
  \bibinfo {author} {\bibfnamefont {F.}~\bibnamefont {Pan}},\ }\href
  {https://doi.org/10.1038/s41578-025-00779-1} {\bibfield  {journal} {\bibinfo
  {journal} {Nature Reviews Materials}\ }\textbf {\bibinfo {volume} {10}},\
  \bibinfo {pages} {473} (\bibinfo {year} {2025})}\BibitemShut {NoStop}%
\bibitem [{\citenamefont {Jungwirth}\ \emph {et~al.}(2025)\citenamefont
  {Jungwirth}, \citenamefont {Fernandes}, \citenamefont {Fradkin},
  \citenamefont {MacDonald}, \citenamefont {Sinova},\ and\ \citenamefont
  {{\v{S}}mejkal}}]{Jungwirth2025Newton}%
  \BibitemOpen
  \bibfield  {author} {\bibinfo {author} {\bibfnamefont {T.}~\bibnamefont
  {Jungwirth}}, \bibinfo {author} {\bibfnamefont {R.~M.}\ \bibnamefont
  {Fernandes}}, \bibinfo {author} {\bibfnamefont {E.}~\bibnamefont {Fradkin}},
  \bibinfo {author} {\bibfnamefont {A.~H.}\ \bibnamefont {MacDonald}}, \bibinfo
  {author} {\bibfnamefont {J.}~\bibnamefont {Sinova}},\ and\ \bibinfo {author}
  {\bibfnamefont {L.}~\bibnamefont {{\v{S}}mejkal}},\ }\bibfield  {journal}
  {\bibinfo  {journal} {Newton}\ }\textbf {\bibinfo {volume} {1}},\ \href
  {https://doi.org/10.1016/j.newton.2025.100162} {10.1016/j.newton.2025.100162}
  (\bibinfo {year} {2025})\BibitemShut {NoStop}%
\bibitem [{\citenamefont {Liu}\ \emph {et~al.}(2022)\citenamefont {Liu},
  \citenamefont {Li}, \citenamefont {Han}, \citenamefont {Wan},\ and\
  \citenamefont {Liu}}]{LiuPF2022PRX}%
  \BibitemOpen
  \bibfield  {author} {\bibinfo {author} {\bibfnamefont {P.}~\bibnamefont
  {Liu}}, \bibinfo {author} {\bibfnamefont {J.}~\bibnamefont {Li}}, \bibinfo
  {author} {\bibfnamefont {J.}~\bibnamefont {Han}}, \bibinfo {author}
  {\bibfnamefont {X.}~\bibnamefont {Wan}},\ and\ \bibinfo {author}
  {\bibfnamefont {Q.}~\bibnamefont {Liu}},\ }\href
  {https://doi.org/10.1103/PhysRevX.12.021016} {\bibfield  {journal} {\bibinfo
  {journal} {Phys. Rev. X}\ }\textbf {\bibinfo {volume} {12}},\ \bibinfo
  {pages} {021016} (\bibinfo {year} {2022})}\BibitemShut {NoStop}%
\bibitem [{\citenamefont {Xiao}\ \emph {et~al.}(2024)\citenamefont {Xiao},
  \citenamefont {Zhao}, \citenamefont {Li}, \citenamefont {Shindou},\ and\
  \citenamefont {Song}}]{XiaoZY2024PRX}%
  \BibitemOpen
  \bibfield  {author} {\bibinfo {author} {\bibfnamefont {Z.}~\bibnamefont
  {Xiao}}, \bibinfo {author} {\bibfnamefont {J.}~\bibnamefont {Zhao}}, \bibinfo
  {author} {\bibfnamefont {Y.}~\bibnamefont {Li}}, \bibinfo {author}
  {\bibfnamefont {R.}~\bibnamefont {Shindou}},\ and\ \bibinfo {author}
  {\bibfnamefont {Z.-D.}\ \bibnamefont {Song}},\ }\href
  {https://doi.org/10.1103/PhysRevX.14.031037} {\bibfield  {journal} {\bibinfo
  {journal} {Phys. Rev. X}\ }\textbf {\bibinfo {volume} {14}},\ \bibinfo
  {pages} {031037} (\bibinfo {year} {2024})}\BibitemShut {NoStop}%
\bibitem [{\citenamefont {Xiao}\ \emph {et~al.}(2026)\citenamefont {Xiao},
  \citenamefont {Li}, \citenamefont {Han}, \citenamefont {Gan}, \citenamefont
  {Yang}, \citenamefont {Shao}, \citenamefont {Zhang}, \citenamefont {Gao},
  \citenamefont {Tian},\ and\ \citenamefont {Zhou}}]{XiaoRC2026SCPMA}%
  \BibitemOpen
  \bibfield  {author} {\bibinfo {author} {\bibfnamefont {R.-C.}\ \bibnamefont
  {Xiao}}, \bibinfo {author} {\bibfnamefont {H.}~\bibnamefont {Li}}, \bibinfo
  {author} {\bibfnamefont {H.}~\bibnamefont {Han}}, \bibinfo {author}
  {\bibfnamefont {W.}~\bibnamefont {Gan}}, \bibinfo {author} {\bibfnamefont
  {M.}~\bibnamefont {Yang}}, \bibinfo {author} {\bibfnamefont {D.-F.}\
  \bibnamefont {Shao}}, \bibinfo {author} {\bibfnamefont {S.-H.}\ \bibnamefont
  {Zhang}}, \bibinfo {author} {\bibfnamefont {Y.}~\bibnamefont {Gao}}, \bibinfo
  {author} {\bibfnamefont {M.}~\bibnamefont {Tian}},\ and\ \bibinfo {author}
  {\bibfnamefont {J.}~\bibnamefont {Zhou}},\ }\href
  {https://doi.org/https://doi.org/10.1007/s11433-025-2769-6} {\bibfield
  {journal} {\bibinfo  {journal} {SCIENCE CHINA Physics, Mechanics and
  Astronomy}\ }\textbf {\bibinfo {volume} {69}},\ \bibinfo {pages} {217511}
  (\bibinfo {year} {2026})}\BibitemShut {NoStop}%
\bibitem [{\citenamefont {Bai}\ \emph {et~al.}(2022)\citenamefont {Bai},
  \citenamefont {Han}, \citenamefont {Feng}, \citenamefont {Zhou},
  \citenamefont {Su}, \citenamefont {Wang}, \citenamefont {Liao}, \citenamefont
  {Zhu}, \citenamefont {Chen}, \citenamefont {Pan}, \citenamefont {Fan},\ and\
  \citenamefont {Song}}]{BaiH2022PRL}%
  \BibitemOpen
  \bibfield  {author} {\bibinfo {author} {\bibfnamefont {H.}~\bibnamefont
  {Bai}}, \bibinfo {author} {\bibfnamefont {L.}~\bibnamefont {Han}}, \bibinfo
  {author} {\bibfnamefont {X.~Y.}\ \bibnamefont {Feng}}, \bibinfo {author}
  {\bibfnamefont {Y.~J.}\ \bibnamefont {Zhou}}, \bibinfo {author}
  {\bibfnamefont {R.~X.}\ \bibnamefont {Su}}, \bibinfo {author} {\bibfnamefont
  {Q.}~\bibnamefont {Wang}}, \bibinfo {author} {\bibfnamefont {L.~Y.}\
  \bibnamefont {Liao}}, \bibinfo {author} {\bibfnamefont {W.~X.}\ \bibnamefont
  {Zhu}}, \bibinfo {author} {\bibfnamefont {X.~Z.}\ \bibnamefont {Chen}},
  \bibinfo {author} {\bibfnamefont {F.}~\bibnamefont {Pan}}, \bibinfo {author}
  {\bibfnamefont {X.~L.}\ \bibnamefont {Fan}},\ and\ \bibinfo {author}
  {\bibfnamefont {C.}~\bibnamefont {Song}},\ }\href
  {https://doi.org/10.1103/PhysRevLett.128.197202} {\bibfield  {journal}
  {\bibinfo  {journal} {Phys. Rev. Lett.}\ }\textbf {\bibinfo {volume} {128}},\
  \bibinfo {pages} {197202} (\bibinfo {year} {2022})}\BibitemShut {NoStop}%
\bibitem [{\citenamefont {Hiraishi}\ \emph {et~al.}(2024)\citenamefont
  {Hiraishi}, \citenamefont {Okabe}, \citenamefont {Koda}, \citenamefont
  {Kadono}, \citenamefont {Muroi}, \citenamefont {Hirai},\ and\ \citenamefont
  {Hiroi}}]{Hiraishi2024PRL}%
  \BibitemOpen
  \bibfield  {author} {\bibinfo {author} {\bibfnamefont {M.}~\bibnamefont
  {Hiraishi}}, \bibinfo {author} {\bibfnamefont {H.}~\bibnamefont {Okabe}},
  \bibinfo {author} {\bibfnamefont {A.}~\bibnamefont {Koda}}, \bibinfo {author}
  {\bibfnamefont {R.}~\bibnamefont {Kadono}}, \bibinfo {author} {\bibfnamefont
  {T.}~\bibnamefont {Muroi}}, \bibinfo {author} {\bibfnamefont
  {D.}~\bibnamefont {Hirai}},\ and\ \bibinfo {author} {\bibfnamefont
  {Z.}~\bibnamefont {Hiroi}},\ }\href
  {https://doi.org/10.1103/PhysRevLett.132.166702} {\bibfield  {journal}
  {\bibinfo  {journal} {Phys. Rev. Lett.}\ }\textbf {\bibinfo {volume} {132}},\
  \bibinfo {pages} {166702} (\bibinfo {year} {2024})}\BibitemShut {NoStop}%
\bibitem [{\citenamefont {Fedchenko}\ \emph {et~al.}(2024)\citenamefont
  {Fedchenko}, \citenamefont {Minár}, \citenamefont {Akashdeep}, \citenamefont
  {D’Souza}, \citenamefont {Vasilyev}, \citenamefont {Tkach}, \citenamefont
  {Odenbreit}, \citenamefont {Nguyen}, \citenamefont {Kutnyakhov},
  \citenamefont {Wind}, \citenamefont {Wenthaus}, \citenamefont {Scholz},
  \citenamefont {Rossnagel}, \citenamefont {Hoesch}, \citenamefont
  {Aeschlimann}, \citenamefont {Stadtmüller}, \citenamefont {Kläui},
  \citenamefont {Schönhense}, \citenamefont {Jungwirth}, \citenamefont
  {Hellenes}, \citenamefont {Jakob}, \citenamefont {Šmejkal}, \citenamefont
  {Sinova},\ and\ \citenamefont {Elmers}}]{Fedchenko2024SA}%
  \BibitemOpen
  \bibfield  {author} {\bibinfo {author} {\bibfnamefont {O.}~\bibnamefont
  {Fedchenko}}, \bibinfo {author} {\bibfnamefont {J.}~\bibnamefont {Minár}},
  \bibinfo {author} {\bibfnamefont {A.}~\bibnamefont {Akashdeep}}, \bibinfo
  {author} {\bibfnamefont {S.~W.}\ \bibnamefont {D’Souza}}, \bibinfo {author}
  {\bibfnamefont {D.}~\bibnamefont {Vasilyev}}, \bibinfo {author}
  {\bibfnamefont {O.}~\bibnamefont {Tkach}}, \bibinfo {author} {\bibfnamefont
  {L.}~\bibnamefont {Odenbreit}}, \bibinfo {author} {\bibfnamefont
  {Q.}~\bibnamefont {Nguyen}}, \bibinfo {author} {\bibfnamefont
  {D.}~\bibnamefont {Kutnyakhov}}, \bibinfo {author} {\bibfnamefont
  {N.}~\bibnamefont {Wind}}, \bibinfo {author} {\bibfnamefont {L.}~\bibnamefont
  {Wenthaus}}, \bibinfo {author} {\bibfnamefont {M.}~\bibnamefont {Scholz}},
  \bibinfo {author} {\bibfnamefont {K.}~\bibnamefont {Rossnagel}}, \bibinfo
  {author} {\bibfnamefont {M.}~\bibnamefont {Hoesch}}, \bibinfo {author}
  {\bibfnamefont {M.}~\bibnamefont {Aeschlimann}}, \bibinfo {author}
  {\bibfnamefont {B.}~\bibnamefont {Stadtmüller}}, \bibinfo {author}
  {\bibfnamefont {M.}~\bibnamefont {Kläui}}, \bibinfo {author} {\bibfnamefont
  {G.}~\bibnamefont {Schönhense}}, \bibinfo {author} {\bibfnamefont
  {T.}~\bibnamefont {Jungwirth}}, \bibinfo {author} {\bibfnamefont {A.~B.}\
  \bibnamefont {Hellenes}}, \bibinfo {author} {\bibfnamefont {G.}~\bibnamefont
  {Jakob}}, \bibinfo {author} {\bibfnamefont {L.}~\bibnamefont {Šmejkal}},
  \bibinfo {author} {\bibfnamefont {J.}~\bibnamefont {Sinova}},\ and\ \bibinfo
  {author} {\bibfnamefont {H.-J.}\ \bibnamefont {Elmers}},\ }\href
  {https://doi.org/10.1126/sciadv.adj4883} {\bibfield  {journal} {\bibinfo
  {journal} {Science Advances}\ }\textbf {\bibinfo {volume} {10}},\ \bibinfo
  {pages} {eadj4883} (\bibinfo {year} {2024})}\BibitemShut {NoStop}%
\bibitem [{\citenamefont {Chen}\ \emph
  {et~al.}(2026{\natexlab{b}})\citenamefont {Chen}, \citenamefont {Qin},
  \citenamefont {Meng}, \citenamefont {Zhao}, \citenamefont {Chen},
  \citenamefont {Xi}, \citenamefont {Wang}, \citenamefont {Liu}, \citenamefont
  {Duan}, \citenamefont {Jiang}, \citenamefont {Li}, \citenamefont {Tan},
  \citenamefont {Liu}, \citenamefont {Wang}, \citenamefont {Liu}, \citenamefont
  {Jiang},\ and\ \citenamefont {Liu}}]{Chen2026NN}%
  \BibitemOpen
  \bibfield  {author} {\bibinfo {author} {\bibfnamefont {H.}~\bibnamefont
  {Chen}}, \bibinfo {author} {\bibfnamefont {P.}~\bibnamefont {Qin}}, \bibinfo
  {author} {\bibfnamefont {Z.}~\bibnamefont {Meng}}, \bibinfo {author}
  {\bibfnamefont {G.}~\bibnamefont {Zhao}}, \bibinfo {author} {\bibfnamefont
  {K.}~\bibnamefont {Chen}}, \bibinfo {author} {\bibfnamefont {C.}~\bibnamefont
  {Xi}}, \bibinfo {author} {\bibfnamefont {X.}~\bibnamefont {Wang}}, \bibinfo
  {author} {\bibfnamefont {L.}~\bibnamefont {Liu}}, \bibinfo {author}
  {\bibfnamefont {Z.}~\bibnamefont {Duan}}, \bibinfo {author} {\bibfnamefont
  {S.}~\bibnamefont {Jiang}}, \bibinfo {author} {\bibfnamefont
  {J.}~\bibnamefont {Li}}, \bibinfo {author} {\bibfnamefont {X.}~\bibnamefont
  {Tan}}, \bibinfo {author} {\bibfnamefont {J.}~\bibnamefont {Liu}}, \bibinfo
  {author} {\bibfnamefont {J.}~\bibnamefont {Wang}}, \bibinfo {author}
  {\bibfnamefont {H.}~\bibnamefont {Liu}}, \bibinfo {author} {\bibfnamefont
  {C.}~\bibnamefont {Jiang}},\ and\ \bibinfo {author} {\bibfnamefont
  {Z.}~\bibnamefont {Liu}},\ }\bibfield  {journal} {\bibinfo  {journal} {Nature
  Nanotechnology}\ }\href {https://doi.org/10.1038/s41565-026-02159-4}
  {10.1038/s41565-026-02159-4} (\bibinfo {year}
  {2026}{\natexlab{b}})\BibitemShut {NoStop}%
\bibitem [{\citenamefont {Zhou}\ \emph {et~al.}(2024)\citenamefont {Zhou},
  \citenamefont {Feng}, \citenamefont {Zhang}, \citenamefont
  {\ifmmode~\check{S}\else \v{S}\fi{}mejkal}, \citenamefont {Sinova},
  \citenamefont {Mokrousov},\ and\ \citenamefont {Yao}}]{ZhouXD2024PRL}%
  \BibitemOpen
  \bibfield  {author} {\bibinfo {author} {\bibfnamefont {X.}~\bibnamefont
  {Zhou}}, \bibinfo {author} {\bibfnamefont {W.}~\bibnamefont {Feng}}, \bibinfo
  {author} {\bibfnamefont {R.-W.}\ \bibnamefont {Zhang}}, \bibinfo {author}
  {\bibfnamefont {L.}~\bibnamefont {\ifmmode~\check{S}\else \v{S}\fi{}mejkal}},
  \bibinfo {author} {\bibfnamefont {J.}~\bibnamefont {Sinova}}, \bibinfo
  {author} {\bibfnamefont {Y.}~\bibnamefont {Mokrousov}},\ and\ \bibinfo
  {author} {\bibfnamefont {Y.}~\bibnamefont {Yao}},\ }\href
  {https://doi.org/10.1103/PhysRevLett.132.056701} {\bibfield  {journal}
  {\bibinfo  {journal} {Phys. Rev. Lett.}\ }\textbf {\bibinfo {volume} {132}},\
  \bibinfo {pages} {056701} (\bibinfo {year} {2024})}\BibitemShut {NoStop}%
\bibitem [{\citenamefont {Bhowal}\ and\ \citenamefont
  {Spaldin}(2024)}]{Bhowal2024PRX}%
  \BibitemOpen
  \bibfield  {author} {\bibinfo {author} {\bibfnamefont {S.}~\bibnamefont
  {Bhowal}}\ and\ \bibinfo {author} {\bibfnamefont {N.~A.}\ \bibnamefont
  {Spaldin}},\ }\href {https://doi.org/10.1103/PhysRevX.14.011019} {\bibfield
  {journal} {\bibinfo  {journal} {Phys. Rev. X}\ }\textbf {\bibinfo {volume}
  {14}},\ \bibinfo {pages} {011019} (\bibinfo {year} {2024})}\BibitemShut
  {NoStop}%
\bibitem [{\citenamefont {Reichlova}\ \emph {et~al.}(2024)\citenamefont
  {Reichlova}, \citenamefont {Lopes~Seeger}, \citenamefont
  {Gonz{\'a}lez-Hern{\'a}ndez}, \citenamefont {Kounta}, \citenamefont
  {Schlitz}, \citenamefont {Kriegner}, \citenamefont {Ritzinger}, \citenamefont
  {Lammel}, \citenamefont {Leivisk{\"a}}, \citenamefont {Birk~Hellenes},
  \citenamefont {Olejn{\'i}k}, \citenamefont {Pet{\v{r}}i{\v{c}}ek},
  \citenamefont {Dole{\v{z}}al}, \citenamefont {Horak}, \citenamefont
  {Schmoranzerova}, \citenamefont {Badura}, \citenamefont {Bertaina},
  \citenamefont {Thomas}, \citenamefont {Baltz}, \citenamefont {Michez},
  \citenamefont {Sinova}, \citenamefont {Goennenwein}, \citenamefont
  {Jungwirth},\ and\ \citenamefont {{\v{S}}mejkal}}]{Reichlova2024NC}%
  \BibitemOpen
  \bibfield  {author} {\bibinfo {author} {\bibfnamefont {H.}~\bibnamefont
  {Reichlova}}, \bibinfo {author} {\bibfnamefont {R.}~\bibnamefont
  {Lopes~Seeger}}, \bibinfo {author} {\bibfnamefont {R.}~\bibnamefont
  {Gonz{\'a}lez-Hern{\'a}ndez}}, \bibinfo {author} {\bibfnamefont
  {I.}~\bibnamefont {Kounta}}, \bibinfo {author} {\bibfnamefont
  {R.}~\bibnamefont {Schlitz}}, \bibinfo {author} {\bibfnamefont
  {D.}~\bibnamefont {Kriegner}}, \bibinfo {author} {\bibfnamefont
  {P.}~\bibnamefont {Ritzinger}}, \bibinfo {author} {\bibfnamefont
  {M.}~\bibnamefont {Lammel}}, \bibinfo {author} {\bibfnamefont
  {M.}~\bibnamefont {Leivisk{\"a}}}, \bibinfo {author} {\bibfnamefont
  {A.}~\bibnamefont {Birk~Hellenes}}, \bibinfo {author} {\bibfnamefont
  {K.}~\bibnamefont {Olejn{\'i}k}}, \bibinfo {author} {\bibfnamefont
  {V.}~\bibnamefont {Pet{\v{r}}i{\v{c}}ek}}, \bibinfo {author} {\bibfnamefont
  {P.}~\bibnamefont {Dole{\v{z}}al}}, \bibinfo {author} {\bibfnamefont
  {L.}~\bibnamefont {Horak}}, \bibinfo {author} {\bibfnamefont
  {E.}~\bibnamefont {Schmoranzerova}}, \bibinfo {author} {\bibfnamefont
  {A.}~\bibnamefont {Badura}}, \bibinfo {author} {\bibfnamefont
  {S.}~\bibnamefont {Bertaina}}, \bibinfo {author} {\bibfnamefont
  {A.}~\bibnamefont {Thomas}}, \bibinfo {author} {\bibfnamefont
  {V.}~\bibnamefont {Baltz}}, \bibinfo {author} {\bibfnamefont
  {L.}~\bibnamefont {Michez}}, \bibinfo {author} {\bibfnamefont
  {J.}~\bibnamefont {Sinova}}, \bibinfo {author} {\bibfnamefont {S.~T.~B.}\
  \bibnamefont {Goennenwein}}, \bibinfo {author} {\bibfnamefont
  {T.}~\bibnamefont {Jungwirth}},\ and\ \bibinfo {author} {\bibfnamefont
  {L.}~\bibnamefont {{\v{S}}mejkal}},\ }\href
  {https://doi.org/10.1038/s41467-024-48493-w} {\bibfield  {journal} {\bibinfo
  {journal} {Nature Communications}\ }\textbf {\bibinfo {volume} {15}},\
  \bibinfo {pages} {4961} (\bibinfo {year} {2024})}\BibitemShut {NoStop}%
\bibitem [{\citenamefont {Jiang}\ \emph {et~al.}(2025)\citenamefont {Jiang},
  \citenamefont {Hu}, \citenamefont {Bai}, \citenamefont {Song}, \citenamefont
  {Mu}, \citenamefont {Qu}, \citenamefont {Li}, \citenamefont {Zhu},
  \citenamefont {Pi}, \citenamefont {Wei}, \citenamefont {Sun}, \citenamefont
  {Huang}, \citenamefont {Zheng}, \citenamefont {Peng}, \citenamefont {He},
  \citenamefont {Li}, \citenamefont {Luo}, \citenamefont {Li}, \citenamefont
  {Chen}, \citenamefont {Li}, \citenamefont {Weng},\ and\ \citenamefont
  {Qian}}]{JiangB2025NP}%
  \BibitemOpen
  \bibfield  {author} {\bibinfo {author} {\bibfnamefont {B.}~\bibnamefont
  {Jiang}}, \bibinfo {author} {\bibfnamefont {M.}~\bibnamefont {Hu}}, \bibinfo
  {author} {\bibfnamefont {J.}~\bibnamefont {Bai}}, \bibinfo {author}
  {\bibfnamefont {Z.}~\bibnamefont {Song}}, \bibinfo {author} {\bibfnamefont
  {C.}~\bibnamefont {Mu}}, \bibinfo {author} {\bibfnamefont {G.}~\bibnamefont
  {Qu}}, \bibinfo {author} {\bibfnamefont {W.}~\bibnamefont {Li}}, \bibinfo
  {author} {\bibfnamefont {W.}~\bibnamefont {Zhu}}, \bibinfo {author}
  {\bibfnamefont {H.}~\bibnamefont {Pi}}, \bibinfo {author} {\bibfnamefont
  {Z.}~\bibnamefont {Wei}}, \bibinfo {author} {\bibfnamefont {Y.-J.}\
  \bibnamefont {Sun}}, \bibinfo {author} {\bibfnamefont {Y.}~\bibnamefont
  {Huang}}, \bibinfo {author} {\bibfnamefont {X.}~\bibnamefont {Zheng}},
  \bibinfo {author} {\bibfnamefont {Y.}~\bibnamefont {Peng}}, \bibinfo {author}
  {\bibfnamefont {L.}~\bibnamefont {He}}, \bibinfo {author} {\bibfnamefont
  {S.}~\bibnamefont {Li}}, \bibinfo {author} {\bibfnamefont {J.}~\bibnamefont
  {Luo}}, \bibinfo {author} {\bibfnamefont {Z.}~\bibnamefont {Li}}, \bibinfo
  {author} {\bibfnamefont {G.}~\bibnamefont {Chen}}, \bibinfo {author}
  {\bibfnamefont {H.}~\bibnamefont {Li}}, \bibinfo {author} {\bibfnamefont
  {H.}~\bibnamefont {Weng}},\ and\ \bibinfo {author} {\bibfnamefont
  {T.}~\bibnamefont {Qian}},\ }\href
  {https://doi.org/10.1038/s41567-025-02822-y} {\bibfield  {journal} {\bibinfo
  {journal} {Nature Physics}\ }\textbf {\bibinfo {volume} {21}},\ \bibinfo
  {pages} {754} (\bibinfo {year} {2025})}\BibitemShut {NoStop}%
\bibitem [{\citenamefont {Chen}\ \emph {et~al.}(2014)\citenamefont {Chen},
  \citenamefont {Niu},\ and\ \citenamefont {MacDonald}}]{ChenH2014PRL}%
  \BibitemOpen
  \bibfield  {author} {\bibinfo {author} {\bibfnamefont {H.}~\bibnamefont
  {Chen}}, \bibinfo {author} {\bibfnamefont {Q.}~\bibnamefont {Niu}},\ and\
  \bibinfo {author} {\bibfnamefont {A.~H.}\ \bibnamefont {MacDonald}},\ }\href
  {https://doi.org/10.1103/PhysRevLett.112.017205} {\bibfield  {journal}
  {\bibinfo  {journal} {Phys. Rev. Lett.}\ }\textbf {\bibinfo {volume} {112}},\
  \bibinfo {pages} {017205} (\bibinfo {year} {2014})}\BibitemShut {NoStop}%
\bibitem [{\citenamefont {Nakatsuji}\ \emph {et~al.}(2015)\citenamefont
  {Nakatsuji}, \citenamefont {Kiyohara},\ and\ \citenamefont
  {Higo}}]{Nakatsuji2015Nature}%
  \BibitemOpen
  \bibfield  {author} {\bibinfo {author} {\bibfnamefont {S.}~\bibnamefont
  {Nakatsuji}}, \bibinfo {author} {\bibfnamefont {N.}~\bibnamefont
  {Kiyohara}},\ and\ \bibinfo {author} {\bibfnamefont {T.}~\bibnamefont
  {Higo}},\ }\href {https://doi.org/10.1038/nature15723} {\bibfield  {journal}
  {\bibinfo  {journal} {Nature}\ }\textbf {\bibinfo {volume} {527}},\ \bibinfo
  {pages} {212} (\bibinfo {year} {2015})}\BibitemShut {NoStop}%
\bibitem [{\citenamefont {Liu}\ and\ \citenamefont
  {Balents}(2017)}]{LiuJP2017PRL}%
  \BibitemOpen
  \bibfield  {author} {\bibinfo {author} {\bibfnamefont {J.}~\bibnamefont
  {Liu}}\ and\ \bibinfo {author} {\bibfnamefont {L.}~\bibnamefont {Balents}},\
  }\href {https://doi.org/10.1103/PhysRevLett.119.087202} {\bibfield  {journal}
  {\bibinfo  {journal} {Phys. Rev. Lett.}\ }\textbf {\bibinfo {volume} {119}},\
  \bibinfo {pages} {087202} (\bibinfo {year} {2017})}\BibitemShut {NoStop}%
\bibitem [{\citenamefont {Li}\ \emph {et~al.}(2017)\citenamefont {Li},
  \citenamefont {Xu}, \citenamefont {Ding}, \citenamefont {Wang}, \citenamefont
  {Shen}, \citenamefont {Lu}, \citenamefont {Zhu},\ and\ \citenamefont
  {Behnia}}]{LiXK2017PRL}%
  \BibitemOpen
  \bibfield  {author} {\bibinfo {author} {\bibfnamefont {X.}~\bibnamefont
  {Li}}, \bibinfo {author} {\bibfnamefont {L.}~\bibnamefont {Xu}}, \bibinfo
  {author} {\bibfnamefont {L.}~\bibnamefont {Ding}}, \bibinfo {author}
  {\bibfnamefont {J.}~\bibnamefont {Wang}}, \bibinfo {author} {\bibfnamefont
  {M.}~\bibnamefont {Shen}}, \bibinfo {author} {\bibfnamefont {X.}~\bibnamefont
  {Lu}}, \bibinfo {author} {\bibfnamefont {Z.}~\bibnamefont {Zhu}},\ and\
  \bibinfo {author} {\bibfnamefont {K.}~\bibnamefont {Behnia}},\ }\href
  {https://doi.org/10.1103/PhysRevLett.119.056601} {\bibfield  {journal}
  {\bibinfo  {journal} {Phys. Rev. Lett.}\ }\textbf {\bibinfo {volume} {119}},\
  \bibinfo {pages} {056601} (\bibinfo {year} {2017})}\BibitemShut {NoStop}%
\bibitem [{\citenamefont {Kimata}\ \emph {et~al.}(2019)\citenamefont {Kimata},
  \citenamefont {Chen}, \citenamefont {Kondou}, \citenamefont {Sugimoto},
  \citenamefont {Muduli}, \citenamefont {Ikhlas}, \citenamefont {Omori},
  \citenamefont {Tomita}, \citenamefont {MacDonald}, \citenamefont
  {Nakatsuji},\ and\ \citenamefont {Otani}}]{Kimata2019Nature}%
  \BibitemOpen
  \bibfield  {author} {\bibinfo {author} {\bibfnamefont {M.}~\bibnamefont
  {Kimata}}, \bibinfo {author} {\bibfnamefont {H.}~\bibnamefont {Chen}},
  \bibinfo {author} {\bibfnamefont {K.}~\bibnamefont {Kondou}}, \bibinfo
  {author} {\bibfnamefont {S.}~\bibnamefont {Sugimoto}}, \bibinfo {author}
  {\bibfnamefont {P.~K.}\ \bibnamefont {Muduli}}, \bibinfo {author}
  {\bibfnamefont {M.}~\bibnamefont {Ikhlas}}, \bibinfo {author} {\bibfnamefont
  {Y.}~\bibnamefont {Omori}}, \bibinfo {author} {\bibfnamefont
  {T.}~\bibnamefont {Tomita}}, \bibinfo {author} {\bibfnamefont {A.~H.}\
  \bibnamefont {MacDonald}}, \bibinfo {author} {\bibfnamefont {S.}~\bibnamefont
  {Nakatsuji}},\ and\ \bibinfo {author} {\bibfnamefont {Y.}~\bibnamefont
  {Otani}},\ }\href {https://doi.org/10.1038/s41586-018-0853-0} {\bibfield
  {journal} {\bibinfo  {journal} {Nature}\ }\textbf {\bibinfo {volume} {565}},\
  \bibinfo {pages} {627} (\bibinfo {year} {2019})}\BibitemShut {NoStop}%
\bibitem [{\citenamefont {Tsai}\ \emph {et~al.}(2020)\citenamefont {Tsai},
  \citenamefont {Higo}, \citenamefont {Kondou}, \citenamefont {Nomoto},
  \citenamefont {Sakai}, \citenamefont {Kobayashi}, \citenamefont {Nakano},
  \citenamefont {Yakushiji}, \citenamefont {Arita}, \citenamefont {Miwa},
  \citenamefont {Otani},\ and\ \citenamefont {Nakatsuji}}]{Tsai2020Nature}%
  \BibitemOpen
  \bibfield  {author} {\bibinfo {author} {\bibfnamefont {H.}~\bibnamefont
  {Tsai}}, \bibinfo {author} {\bibfnamefont {T.}~\bibnamefont {Higo}}, \bibinfo
  {author} {\bibfnamefont {K.}~\bibnamefont {Kondou}}, \bibinfo {author}
  {\bibfnamefont {T.}~\bibnamefont {Nomoto}}, \bibinfo {author} {\bibfnamefont
  {A.}~\bibnamefont {Sakai}}, \bibinfo {author} {\bibfnamefont
  {A.}~\bibnamefont {Kobayashi}}, \bibinfo {author} {\bibfnamefont
  {T.}~\bibnamefont {Nakano}}, \bibinfo {author} {\bibfnamefont
  {K.}~\bibnamefont {Yakushiji}}, \bibinfo {author} {\bibfnamefont
  {R.}~\bibnamefont {Arita}}, \bibinfo {author} {\bibfnamefont
  {S.}~\bibnamefont {Miwa}}, \bibinfo {author} {\bibfnamefont {Y.}~\bibnamefont
  {Otani}},\ and\ \bibinfo {author} {\bibfnamefont {S.}~\bibnamefont
  {Nakatsuji}},\ }\href {https://doi.org/10.1038/s41586-020-2211-2} {\bibfield
  {journal} {\bibinfo  {journal} {Nature}\ }\textbf {\bibinfo {volume} {580}},\
  \bibinfo {pages} {608} (\bibinfo {year} {2020})}\BibitemShut {NoStop}%
\bibitem [{\citenamefont {Go}\ \emph {et~al.}(2022)\citenamefont {Go},
  \citenamefont {Sallermann}, \citenamefont {Lux}, \citenamefont {Bl\"ugel},
  \citenamefont {Gomonay},\ and\ \citenamefont {Mokrousov}}]{Go2022PRL}%
  \BibitemOpen
  \bibfield  {author} {\bibinfo {author} {\bibfnamefont {D.}~\bibnamefont
  {Go}}, \bibinfo {author} {\bibfnamefont {M.}~\bibnamefont {Sallermann}},
  \bibinfo {author} {\bibfnamefont {F.~R.}\ \bibnamefont {Lux}}, \bibinfo
  {author} {\bibfnamefont {S.}~\bibnamefont {Bl\"ugel}}, \bibinfo {author}
  {\bibfnamefont {O.}~\bibnamefont {Gomonay}},\ and\ \bibinfo {author}
  {\bibfnamefont {Y.}~\bibnamefont {Mokrousov}},\ }\href
  {https://doi.org/10.1103/PhysRevLett.129.097204} {\bibfield  {journal}
  {\bibinfo  {journal} {Phys. Rev. Lett.}\ }\textbf {\bibinfo {volume} {129}},\
  \bibinfo {pages} {097204} (\bibinfo {year} {2022})}\BibitemShut {NoStop}%
\bibitem [{\citenamefont {Tsai}\ \emph {et~al.}(2026)\citenamefont {Tsai},
  \citenamefont {Matsuda}, \citenamefont {Kondou}, \citenamefont {Shimizu},
  \citenamefont {Nomoto}, \citenamefont {Higo}, \citenamefont {Matsuo},
  \citenamefont {Tsushima}, \citenamefont {Asakura}, \citenamefont {Peng},
  \citenamefont {Nishio-Hamane}, \citenamefont {Yamada}, \citenamefont {Tang},
  \citenamefont {Iizuka}, \citenamefont {Miwa}, \citenamefont {Arita},
  \citenamefont {Takenaka},\ and\ \citenamefont {Nakatsuji}}]{Tsai2026Science}%
  \BibitemOpen
  \bibfield  {author} {\bibinfo {author} {\bibfnamefont {H.}~\bibnamefont
  {Tsai}}, \bibinfo {author} {\bibfnamefont {T.}~\bibnamefont {Matsuda}},
  \bibinfo {author} {\bibfnamefont {K.}~\bibnamefont {Kondou}}, \bibinfo
  {author} {\bibfnamefont {K.}~\bibnamefont {Shimizu}}, \bibinfo {author}
  {\bibfnamefont {T.}~\bibnamefont {Nomoto}}, \bibinfo {author} {\bibfnamefont
  {T.}~\bibnamefont {Higo}}, \bibinfo {author} {\bibfnamefont {T.}~\bibnamefont
  {Matsuo}}, \bibinfo {author} {\bibfnamefont {Y.}~\bibnamefont {Tsushima}},
  \bibinfo {author} {\bibfnamefont {M.}~\bibnamefont {Asakura}}, \bibinfo
  {author} {\bibfnamefont {H.}~\bibnamefont {Peng}}, \bibinfo {author}
  {\bibfnamefont {D.}~\bibnamefont {Nishio-Hamane}}, \bibinfo {author}
  {\bibfnamefont {S.}~\bibnamefont {Yamada}}, \bibinfo {author} {\bibfnamefont
  {R.}~\bibnamefont {Tang}}, \bibinfo {author} {\bibfnamefont {T.}~\bibnamefont
  {Iizuka}}, \bibinfo {author} {\bibfnamefont {S.}~\bibnamefont {Miwa}},
  \bibinfo {author} {\bibfnamefont {R.}~\bibnamefont {Arita}}, \bibinfo
  {author} {\bibfnamefont {M.}~\bibnamefont {Takenaka}},\ and\ \bibinfo
  {author} {\bibfnamefont {S.}~\bibnamefont {Nakatsuji}},\ }\href
  {https://doi.org/10.1126/science.adt3136} {\bibfield  {journal} {\bibinfo
  {journal} {Science}\ }\textbf {\bibinfo {volume} {392}},\ \bibinfo {pages}
  {761} (\bibinfo {year} {2026})}\BibitemShut {NoStop}%
\bibitem [{\citenamefont {Rimmler}\ \emph {et~al.}(2025)\citenamefont
  {Rimmler}, \citenamefont {Pal},\ and\ \citenamefont
  {Parkin}}]{Rimmler2025NRM}%
  \BibitemOpen
  \bibfield  {author} {\bibinfo {author} {\bibfnamefont {B.~H.}\ \bibnamefont
  {Rimmler}}, \bibinfo {author} {\bibfnamefont {B.}~\bibnamefont {Pal}},\ and\
  \bibinfo {author} {\bibfnamefont {S.~S.~P.}\ \bibnamefont {Parkin}},\ }\href
  {https://doi.org/10.1038/s41578-024-00706-w} {\bibfield  {journal} {\bibinfo
  {journal} {Nature Reviews Materials}\ }\textbf {\bibinfo {volume} {10}},\
  \bibinfo {pages} {109} (\bibinfo {year} {2025})}\BibitemShut {NoStop}%
\bibitem [{\citenamefont {Cederholm}\ \emph {et~al.}(2026)\citenamefont
  {Cederholm}, \citenamefont {Xu}, \citenamefont {Guo}, \citenamefont {Ovesen},
  \citenamefont {Olsen}, \citenamefont {Krighaar}, \citenamefont {Knekna},
  \citenamefont {Soh}, \citenamefont {Lee}, \citenamefont {Qureshi},
  \citenamefont {Velamazan}, \citenamefont {Ressouche}, \citenamefont
  {Boothroyd},\ and\ \citenamefont {Jacobsen}}]{cederholm2026arxiv}%
  \BibitemOpen
  \bibfield  {author} {\bibinfo {author} {\bibfnamefont {J.~J.}\ \bibnamefont
  {Cederholm}}, \bibinfo {author} {\bibfnamefont {Z.}~\bibnamefont {Xu}},
  \bibinfo {author} {\bibfnamefont {Y.}~\bibnamefont {Guo}}, \bibinfo {author}
  {\bibfnamefont {M.}~\bibnamefont {Ovesen}}, \bibinfo {author} {\bibfnamefont
  {T.}~\bibnamefont {Olsen}}, \bibinfo {author} {\bibfnamefont {K.~M.~L.}\
  \bibnamefont {Krighaar}}, \bibinfo {author} {\bibfnamefont {C.}~\bibnamefont
  {Knekna}}, \bibinfo {author} {\bibfnamefont {J.~R.}\ \bibnamefont {Soh}},
  \bibinfo {author} {\bibfnamefont {Y.}~\bibnamefont {Lee}}, \bibinfo {author}
  {\bibfnamefont {N.}~\bibnamefont {Qureshi}}, \bibinfo {author} {\bibfnamefont
  {J.~A.~R.}\ \bibnamefont {Velamazan}}, \bibinfo {author} {\bibfnamefont
  {E.}~\bibnamefont {Ressouche}}, \bibinfo {author} {\bibfnamefont {A.~T.}\
  \bibnamefont {Boothroyd}},\ and\ \bibinfo {author} {\bibfnamefont
  {H.}~\bibnamefont {Jacobsen}},\ }\href {https://arxiv.org/abs/2510.06808}
  {\bibfield  {journal} {\bibinfo  {journal} {2510.06808V2}\ } (\bibinfo {year}
  {2026})}\BibitemShut {NoStop}%
\bibitem [{Mn3()}]{Mn3Sn}%
  \BibitemOpen
  \href@noop {} {\bibinfo  {journal} {Both structures permit multiple magnetic
  domains, with complex domain-wall configurations further complicating the
  interpretation of diffraction data. DFT calculations reveal that the energy
  difference between these configurations is vanishingly small--degenerate
  within computational uncertainty~\cite{cederholm2026arxiv}}\ }\BibitemShut
  {NoStop}%
\bibitem [{\citenamefont {Liu}\ \emph {et~al.}(2026)\citenamefont {Liu},
  \citenamefont {Gao},\ and\ \citenamefont {Niu}}]{LiuZ2026PRL}%
  \BibitemOpen
\bibfield  {journal} {  }\bibfield  {author} {\bibinfo {author} {\bibfnamefont
  {Z.}~\bibnamefont {Liu}}, \bibinfo {author} {\bibfnamefont {Y.}~\bibnamefont
  {Gao}},\ and\ \bibinfo {author} {\bibfnamefont {Q.}~\bibnamefont {Niu}},\
  }\href {https://doi.org/10.1103/64wt-51gd} {\bibfield  {journal} {\bibinfo
  {journal} {Phys. Rev. Lett.}\ }\textbf {\bibinfo {volume} {136}},\ \bibinfo
  {pages} {026703} (\bibinfo {year} {2026})}\BibitemShut {NoStop}%
\bibitem [{\citenamefont {Berdyugin}\ \emph {et~al.}(2019)\citenamefont
  {Berdyugin}, \citenamefont {Xu}, \citenamefont {Pellegrino}, \citenamefont
  {Kumar}, \citenamefont {Principi}, \citenamefont {Torre}, \citenamefont
  {Shalom}, \citenamefont {Taniguchi}, \citenamefont {Watanabe}, \citenamefont
  {Grigorieva}, \citenamefont {Polini}, \citenamefont {Geim},\ and\
  \citenamefont {Bandurin}}]{Berdyugin2019Science}%
  \BibitemOpen
  \bibfield  {author} {\bibinfo {author} {\bibfnamefont {A.~I.}\ \bibnamefont
  {Berdyugin}}, \bibinfo {author} {\bibfnamefont {S.~G.}\ \bibnamefont {Xu}},
  \bibinfo {author} {\bibfnamefont {F.~M.~D.}\ \bibnamefont {Pellegrino}},
  \bibinfo {author} {\bibfnamefont {R.~K.}\ \bibnamefont {Kumar}}, \bibinfo
  {author} {\bibfnamefont {A.}~\bibnamefont {Principi}}, \bibinfo {author}
  {\bibfnamefont {I.}~\bibnamefont {Torre}}, \bibinfo {author} {\bibfnamefont
  {M.~B.}\ \bibnamefont {Shalom}}, \bibinfo {author} {\bibfnamefont
  {T.}~\bibnamefont {Taniguchi}}, \bibinfo {author} {\bibfnamefont
  {K.}~\bibnamefont {Watanabe}}, \bibinfo {author} {\bibfnamefont {I.~V.}\
  \bibnamefont {Grigorieva}}, \bibinfo {author} {\bibfnamefont
  {M.}~\bibnamefont {Polini}}, \bibinfo {author} {\bibfnamefont {A.~K.}\
  \bibnamefont {Geim}},\ and\ \bibinfo {author} {\bibfnamefont {D.~A.}\
  \bibnamefont {Bandurin}},\ }\href {https://doi.org/10.1126/science.aau0685}
  {\bibfield  {journal} {\bibinfo  {journal} {Science}\ }\textbf {\bibinfo
  {volume} {364}},\ \bibinfo {pages} {162} (\bibinfo {year}
  {2019})}\BibitemShut {NoStop}%
\bibitem [{\citenamefont {Levitov}\ and\ \citenamefont
  {Falkovich}(2016)}]{Levitov2016NP}%
  \BibitemOpen
  \bibfield  {author} {\bibinfo {author} {\bibfnamefont {L.}~\bibnamefont
  {Levitov}}\ and\ \bibinfo {author} {\bibfnamefont {G.}~\bibnamefont
  {Falkovich}},\ }\href {https://doi.org/10.1038/nphys3667} {\bibfield
  {journal} {\bibinfo  {journal} {Nature Physics}\ }\textbf {\bibinfo {volume}
  {12}},\ \bibinfo {pages} {672} (\bibinfo {year} {2016})}\BibitemShut
  {NoStop}%
\bibitem [{\citenamefont {Shragai}\ \emph {et~al.}(2026)\citenamefont
  {Shragai}, \citenamefont {Horsley}, \citenamefont {Kim}, \citenamefont
  {Kim},\ and\ \citenamefont {Ramshaw}}]{Shragai2026Nature}%
  \BibitemOpen
  \bibfield  {author} {\bibinfo {author} {\bibfnamefont {A.}~\bibnamefont
  {Shragai}}, \bibinfo {author} {\bibfnamefont {E.}~\bibnamefont {Horsley}},
  \bibinfo {author} {\bibfnamefont {S.}~\bibnamefont {Kim}}, \bibinfo {author}
  {\bibfnamefont {Y.-J.}\ \bibnamefont {Kim}},\ and\ \bibinfo {author}
  {\bibfnamefont {B.~J.}\ \bibnamefont {Ramshaw}},\ }\bibfield  {journal}
  {\bibinfo  {journal} {Nature}\ }\href
  {https://doi.org/10.1038/s41586-026-10420-y} {10.1038/s41586-026-10420-y}
  (\bibinfo {year} {2026})\BibitemShut {NoStop}%
\bibitem [{\citenamefont {Mead}\ and\ \citenamefont
  {Truhlar}(1979)}]{Mead1979JCP}%
  \BibitemOpen
  \bibfield  {author} {\bibinfo {author} {\bibfnamefont {C.~A.}\ \bibnamefont
  {Mead}}\ and\ \bibinfo {author} {\bibfnamefont {D.~G.}\ \bibnamefont
  {Truhlar}},\ }\href {https://doi.org/10.1063/1.437734} {\bibfield  {journal}
  {\bibinfo  {journal} {The Journal of Chemical Physics}\ }\textbf {\bibinfo
  {volume} {70}},\ \bibinfo {pages} {2284} (\bibinfo {year}
  {1979})}\BibitemShut {NoStop}%
\bibitem [{\citenamefont {Qin}\ \emph {et~al.}(2012)\citenamefont {Qin},
  \citenamefont {Zhou},\ and\ \citenamefont {Shi}}]{QinT2012PRB}%
  \BibitemOpen
  \bibfield  {author} {\bibinfo {author} {\bibfnamefont {T.}~\bibnamefont
  {Qin}}, \bibinfo {author} {\bibfnamefont {J.}~\bibnamefont {Zhou}},\ and\
  \bibinfo {author} {\bibfnamefont {J.}~\bibnamefont {Shi}},\ }\href
  {https://doi.org/10.1103/PhysRevB.86.104305} {\bibfield  {journal} {\bibinfo
  {journal} {Phys. Rev. B}\ }\textbf {\bibinfo {volume} {86}},\ \bibinfo
  {pages} {104305} (\bibinfo {year} {2012})}\BibitemShut {NoStop}%
\bibitem [{\citenamefont {Hu}\ \emph {et~al.}(2021)\citenamefont {Hu},
  \citenamefont {Yu}, \citenamefont {Garate},\ and\ \citenamefont
  {Liu}}]{HuLH2021PRL}%
  \BibitemOpen
  \bibfield  {author} {\bibinfo {author} {\bibfnamefont {L.-H.}\ \bibnamefont
  {Hu}}, \bibinfo {author} {\bibfnamefont {J.}~\bibnamefont {Yu}}, \bibinfo
  {author} {\bibfnamefont {I.}~\bibnamefont {Garate}},\ and\ \bibinfo {author}
  {\bibfnamefont {C.-X.}\ \bibnamefont {Liu}},\ }\href
  {https://doi.org/10.1103/PhysRevLett.127.125901} {\bibfield  {journal}
  {\bibinfo  {journal} {Phys. Rev. Lett.}\ }\textbf {\bibinfo {volume} {127}},\
  \bibinfo {pages} {125901} (\bibinfo {year} {2021})}\BibitemShut {NoStop}%
\bibitem [{\citenamefont {Shan}(2022)}]{ShanWY2022PRB}%
  \BibitemOpen
  \bibfield  {author} {\bibinfo {author} {\bibfnamefont {W.-Y.}\ \bibnamefont
  {Shan}},\ }\href {https://doi.org/10.1103/PhysRevB.105.L121302} {\bibfield
  {journal} {\bibinfo  {journal} {Phys. Rev. B}\ }\textbf {\bibinfo {volume}
  {105}},\ \bibinfo {pages} {L121302} (\bibinfo {year} {2022})}\BibitemShut
  {NoStop}%
\bibitem [{\citenamefont {Hu}\ \emph {et~al.}(2025)\citenamefont {Hu},
  \citenamefont {Li}, \citenamefont {Guo}, \citenamefont {Wang},\ and\
  \citenamefont {Chang}}]{HuJM2025PRL}%
  \BibitemOpen
  \bibfield  {author} {\bibinfo {author} {\bibfnamefont {J.}~\bibnamefont
  {Hu}}, \bibinfo {author} {\bibfnamefont {W.}~\bibnamefont {Li}}, \bibinfo
  {author} {\bibfnamefont {Z.}~\bibnamefont {Guo}}, \bibinfo {author}
  {\bibfnamefont {H.}~\bibnamefont {Wang}},\ and\ \bibinfo {author}
  {\bibfnamefont {K.}~\bibnamefont {Chang}},\ }\href
  {https://doi.org/10.1103/y66p-kjj7} {\bibfield  {journal} {\bibinfo
  {journal} {Phys. Rev. Lett.}\ }\textbf {\bibinfo {volume} {135}},\ \bibinfo
  {pages} {256404} (\bibinfo {year} {2025})}\BibitemShut {NoStop}%
\bibitem [{\citenamefont {Riesz}\ and\ \citenamefont
  {Sz.~Nagy}(1955)}]{Riesz1955FA}%
  \BibitemOpen
  \bibfield  {author} {\bibinfo {author} {\bibfnamefont {F.}~\bibnamefont
  {Riesz}}\ and\ \bibinfo {author} {\bibfnamefont {B.}~\bibnamefont
  {Sz.~Nagy}},\ }\href@noop {} {\emph {\bibinfo {title} {Functional
  Analysis}}},\ \bibinfo {edition} {2nd}\ ed.\ (\bibinfo  {publisher} {DOVER
  PUBLICATIONS, INC.},\ \bibinfo {year} {1955})\BibitemShut {NoStop}%
\end{thebibliography}%

\end{document}